\documentclass[lettersize,journal]{IEEEtran}

\usepackage{cite}
\usepackage{amsmath,amssymb,amsfonts,bm}
\usepackage{mathtools}
\usepackage{algorithmic}
\usepackage{multirow}
\usepackage{bigints}
\usepackage{textcomp}
\usepackage{graphicx}
\usepackage[ruled,vlined]{algorithm2e}
\usepackage{textcomp}
\usepackage{algorithmic}
\usepackage{array}
\usepackage{comment}
\usepackage{esvect}
\usepackage{subcaption}
\usepackage{xcolor}
\newcolumntype{M}[1]{>{\centering\arraybackslash}m{#1}}
\newcommand*{\QEDA}{\null\nobreak\hfill\ensuremath{\blacksquare}}%

\newcommand\myeqa{\mathrel{\stackrel{\makebox[0pt]{\mbox{\normalfont\scriptsize (a)}}}{=}}}
\newcommand\myeqb{\mathrel{\stackrel{\makebox[0pt]{\mbox{\normalfont\scriptsize (b)}}}{=}}}
\newcommand\myeqc{\mathrel{\stackrel{\makebox[0pt]{\mbox{\normalfont\scriptsize (c)}}}{=}}}

\newcommand\myeqdef{\mathrel{\stackrel{\makebox[0pt]{\mbox{\normalfont\scriptsize conv}}}{=}}}

\usepackage{stfloats} 

\begin{document}

% paper title
% Titles are generally capitalized except for words such as a, an, and, as,
% at, but, by, for, in, nor, of, on, or, the, to and up, which are usually
% not capitalized unless they are the first or last word of the title.
% Linebreaks \\ can be used within to get better formatting as desired.
% Do not put math or special symbols in the title.
\title{Comprehensive Analysis of Maximum Power Association Policy for Cellular Networks Using Distance and Angular Coordinates}
%  
%
% author names and IEEE memberships
% note positions of commas and nonbreaking spaces ( ~ ) LaTeX will not break
% a structure at a ~ so this keeps an author's name from being broken across
% two lines.
% use \thanks{} to gain access to the first footnote area
% a separate \thanks must be used for each paragraph as LaTeX2e's \thanks
% was not built to handle multiple paragraphs
%

\author{Harris K.~Armeniakos,~\IEEEmembership{Graduate Student Member,~IEEE,} \\
        Athanasios G.~Kanatas,~\IEEEmembership{Senior Member,~IEEE,} \\
        and Harpreet S. Dhillon,~\IEEEmembership{Fellow,~IEEE}% <-this % stops a space
\thanks{H. K. Armeniakos and A. G. Kanatas are with the Department
of Digital Systems, University of Piraeus, 18534 Piraeus, Greece (e-mail:
\{harmen,kanatas\}@unipi.gr).}
\thanks{H. S. Dhillon is with with Wireless@Virginia Tech, Bradley Department of
Electrical and Computer Engineering, Virginia Tech, Blacksburg, 24061, VA, USA (e-mail: hdhillon@vt.edu).}% <-this % stops a space
}

\maketitle

% As a general rule, do not put math, special symbols or citations
% in the abstract or keywords.
\begin{abstract}
A novel stochastic geometry framework is proposed in this paper to study the downlink coverage performance in a millimeter wave (mmWave) cellular network by jointly considering the polar coordinates of the Base Stations (BSs) with respect to the typical user located at the origin. Specifically, both the Euclidean and the angular distances of the BSs in a maximum power-based association policy for the UE are considered to account for realistic beam management considerations, which have been largely ignored in the literature, especially in the cell association phase. For completeness, two other association schemes are considered and exact-form expressions for the coverage probability are derived. Subsequently, the key role of angular distances is highlighted by defining the dominant interferer using angular distance-based criteria instead of Euclidean distance-based, and conducting a dominant interferer-based coverage probability analysis. Among others, the numerical results reveal that considering angular distance-based criteria for determining both the serving and the dominant interfering BS, can approximate the coverage performance more accurately as compared to utilizing Euclidean distance-based criteria. To the best of the authors’ knowledge, this is the first work that rigorously explores the role of angular distances in the association policy and analysis of cellular networks.
\end{abstract}

% Note that keywords are not normally used for peerreview papers.

% For peer review papers, you can put extra information on the cover
% page as needed:
% \ifCLASSOPTIONpeerreview
% \begin{center} \bfseries EDICS Category: 3-BBND \end{center}
% \fi
%
% For peerreview papers, this IEEEtran command inserts a page break and
% creates the second title. It will be ignored for other modes.
\IEEEpeerreviewmaketitle

\section{Introduction}
Due to the rapid proliferation of the smart devices and novel rate-greedy applications, mobile data traffic has witnessed a tremendous growth. As the fifth generation (5G) cellular networks become more ubiquitous and we start thinking of the sixth generation (6G), the  networks will continue to expand to higher frequency bands to address the need for extreme capacity. Recently, millimeter wave (mmWave) and sub-Terahertz (THz) networks, operating at frequencies between 24 and roughly 330 GHz, have attracted considerable attention from both academia and industry due to the enormous available bandwidth \cite{Review1}, \cite{Review2}. However, the extremely high data rates achieved in mmWave and sub-THz bands come with challenging propagation characteristics, with the path loss being a key issue. To overcome these limitations, first, mmWave networks are envisioned to be densely deployed to achieve acceptable coverage \cite{maxpower1}. However, increasing the density of base stations (BSs) leads to severe interference problems which in turn may cause  a significant number of transmission failures. Next, steerable antenna arrays with highly directional antenna beams are needed to  achieve high power gain and improved coverage. Therefore, the antennas' 3 dB beamwidth becomes a key design parameter. In addition, mmWave channels are sensitive to the dynamically changing propagation environment (moving pedestrians, passing cars, blockages, etc.). Thus, 5G New Radio (NR) mmWave wireless networks rely on adaptive beamforming and beam selection techniques for optimizing the performance.   
\par In a network where the user equipment (UEs) and the BSs are equipped with antenna arrays that are able to form directional beams, one of the most rational criteria is to perform the association of a UE to its serving BS based on the maximum received power. Therefore, the probability density function (PDF) of the desired signal received power should be calculated based on both polar coordinates. As we will discuss shortly, if one ignores practical beam management considerations, cell association decisions can be made purely based on the Euclidean distances.  However, in this paper, it is rigorously argued that if one accounts for practical issues such as the limited capability of UEs to perform a perfect beam alignment or an imperfect channel estimation, or a codebook-based beamforming with limited number of beams, then one needs to consider angular distances.

\subsection{Related Work and Motivation}
With the rise of 5G and beyond communication systems, the use of multiple antennas at the BSs and the UEs has introduced beamforming capabilities as a central feature in 5G NR that leads to higher data rates. However, a series of beam management procedures are needed to ensure efficient handling and network operation. The selection of the best receiving beam is performed by measuring the average received signal power in each beam through exhaustive scanning in a set of candidate serving BSs. The maximum power-based association policy is governed by the distance dependent path loss and the transmitting and receiving antenna gain patterns. However, either a binary valued antenna pattern, called flat-top pattern \cite{Chetlur}$- \hspace{-0.15cm}$\cite{mmwave1}, or ideal conditions with realistic patterns, i.e., perfect channel estimation and beam training, that imply full alignment between BS's transmitting and UE's receiving beams has been assumed in most cases. In this \textit{ideal baseline scenario}, the cell association decision is purely based on the Euclidean distance between the two nodes \cite{maxpower1}, \cite{maxpower2}$- \hspace{-0.15cm}$\cite{maxpower7}. 

Many works in the literature have studied beam management techniques and procedures for 5G NR networks by adopting tools from stochastic geometry, since it captures the spatial randomness of network elements \cite{SG1}$- \hspace{-0.15cm}$\cite{SG3}. Modeling the spatial locations of BSs and/or UEs as point processes allows the use of powerful tools from stochastic geometry to derive tractable analytical results for several key performance metrics. Therefore, in \cite{handover}, the authors study among others both the initial beam selection during BS handover and beam reselection technique in a mmWave cell. However, the interference from other BSs is ignored. To address this issue, the authors in \cite{Baccelli} develop a stochastic geometry framework and conduct a detailed performance analysis in terms of the average achievable rate and success probability. Going beyond the coverage probability and the achievable rate, in \cite{TMC}, the authors studied the average number of beam switching and handover events, in mmWave vehicular networks. Beam management techniques were also considered. In \cite{GLOBECOM2021},  both the impact of beamwidth in the reliability and throughput of a THz network and the impact of the highly directional antennas on the beam management procedures was investigated. 
\par In realistic mmWave networks, \textit{beam misalignment} is inevitable, and the direction of the UE's maximum gain may not be necessarily fully aligned with the corresponding one of the serving BS \cite{survey}.  More specifically, beam misalignment can occur between the transmitting and receiving beams after channel estimation during the 5G NR beam management-based UE's association policy \cite{R1} due to the following reasons: 1) use of codebook-based beamforming at the UEs with limited number of beams, 2) imperfect channel estimation, which results in estimation errors in the angle-of-arrival (AoA) or angle-of-departure (AoD), 3) imperfections in the antenna arrays, which includes array perturbation and mutual coupling, 4) mobility of the transceivers, and 5) environmental vibrations such as from wind or moving vehicles. Indeed, by considering codebook-based beamforming at the UE, the receiver is agnostic to the conditions that provide maximum power and the UE will perform scanning within a  distance-limited finite area to select the serving BS among a set of candidate serving BSs. Therefore, it becomes clear that the selection of the serving BS will strongly depend on its  location and thus, one should account for both polar coordinates of the candidate serving BSs in the determination of the maximum receiver power. Therefore, misalignment needs to be carefully accounted for in the mathematical analyses if one is to capture realistic 5G NR beam management-based association procedures. While path-loss is just dependent on the frequency and Euclidean distance, misalignment error, which is now a function of the angular distance between the candidate serving BSs and the UE, should be explicitly considered in the measurement of the received power in the UE's association policy. The authors in \cite{Olson} and \cite{maxpower5} model the misalignment error as a random variable following the truncated Gaussian distribution, whereas the authors in \cite{popovski}, derived an empirical PDF for the misaligned gain based on simulations.  However, the consideration of codebook-based beamforming at the UE necessitates a more nuanced analysis. To the best of the authors' knowledge, this work proposes for the first time a stochastic geometry framework to study the performance in a mmWave cellular network by adopting 5G NR beam management-based procedures and jointly considering the impact of both the Euclidean and angular distances of the BSs in the UE's association policy. The use of the angular distance is critical for the accurate estimation of the receiving antenna gain using  3GPP antenna patterns and allows to depart from the ideal baseline scenario.
\par Along similar lines, many works have captured the effect of the angular coordinate in the calculation of interference power and correspondingly in performance analysis of cellular networks by adopting realistic antenna patterns. In \cite{Banagar}, the authors considered the effect of beam misalignment utilizing a 3GPP-based antenna pattern in a stochastic geometry framework. In \cite{Approx}, the authors investigated the impact of directional antenna arrays on mmWave networks. Among other insights, the role of realistic antenna patterns in the interference power is demonstrated. In \cite{Approx2}, a multi-cosine antenna pattern is proposed to approximate the actual antenna pattern of a uniform linear array (ULA) and the impact in the interference power is highlighted. In \cite{Approx3}, the authors adopt an actual three dimensional antenna model and a uniform planar array, which is mounted on UAVs, to examine the impact of both azimuth and elevation angles on the interference power. 
\par On another front, the idea of the dominant interferer has been introduced in the literature to facilitate the calculation of the SINR and provide a realistic and mathematically tractable approximation of the accurate aggregate interference \cite{dom1}$- \hspace{-0.15cm}$\cite{dom4}. The notion of angular distances and their implication on the identification of the dominant interferer has recently been highlighted in \cite{kanatas} and \cite{globecom}. Accordingly, a dominant interfering BS may not necessarily be the closest one to the receiver. Indeed, a far interferer may cause severer interference than a closer one, due to the fact that the AoA at the receiver may fall within the 3dB beamwidth of the antenna beam.
\par In summary, even though angular coordinates have appeared in the stochastic geometry-based analysis, their manifestation in the received power as a consequence of realistic beam management procedure and the overall effect on the system performance has not been studied, which is the main objective of this paper. Consequently, this work proposes for the first time a stochastic geometry framework to study the implications of beam misalignment error in the association policy and the corresponding performance of a mmWave cellular network by adopting realistic 5G NR beam management-based procedures  and practical equipment limitations. Therefore, extended comparison of the ideal baseline scenario with several association policies is provided, and evaluation of diverse dominant interferer-based scenarios are highlighted.

\subsection{Contributions}
The main contributions of this paper are the following: 

\subsubsection{Angular distance distributions} By exploiting fundamental concepts of stochastic geometry theory, both the $n$th nearest node and joint angular distance distributions within a finite ball are derived in closed form for a two-dimensional (2D) finite homogeneous Poisson point process (HPPP). As it will be shown, these distances play a key role in the definition of dominant interferer in mmWave networks.   

\subsubsection{ Refining the maximum power association policy in a mmWave stochastic geometry framework} By employing the HPPP for modeling the spatial locations of the BSs, a number of BSs with known density are deployed in a mmWave cellular network. The typical UE, which is equipped with a realistic 3GPP-based antenna pattern, exploits directional beamforming capabilities under imperfect alignment to communicate with the serving BS.  In this work, beam misalignment at the UE may occur due to limited codebook-based beamforming at the UE and/or from imperfect channel estimation and/or potentially antenna arrays imperfections \footnote{ According to the beam management procedure of the 3GPP NR for mmWave frequencies \cite{survey}, our framework is  fully consistent with the  Non- Standalone-Uplink (NSA-UL) scheme and beam misalignment can occur during the beam measurement phase as a result of possible array imperfections at the UE and/or imperfect channel estimation.}, and clearly depends on the random spatial location of the serving BS. Accordingly, as a key contribution, a realistic beam management-based association policy that jointly considers both the Euclidean and angular distances of the BSs, is proposed for the UE. The proposed  policy takes into account both the directional beamforming and the beam misalignment error at the UE through the adoption of a 3GPP-based antenna pattern. Two other association schemes, i.e., an ideal baseline scenario and a purely angular distance-based association scheme, are also proposed as special cases for completeness. 

\subsubsection{Coverage probability analysis} Performance analysis in terms of coverage probability is conducted under the three association schemes and analytical  expressions are obtained. As key intermediate results, the PDF of the maximum received power and the Laplace transform (LT) of the aggregate interference power distributions, are derived in exact form.

\subsubsection{Dominant interferer approach} A coverage probability analysis is conducted under the assumption of a single dominant interferer for the maximum received power association policy. As a key result, the joint PDF of the maximum received power from the serving BS and the received power from the dominant interferer is derived. Conventionally, the dominant interferer is considered as the nearest BS based on the Euclidean distance. By extending the definition of the dominant interferer as the nearest BS in angular distance w.r.t. a reference line in mmWave networks, coverage probability analysis is conducted for the two special cases under the assumption of a dominant interferer for which exact-form expressions are derived. To this end,  the PDF of the ratio of the nearest BS, in angular distance, antenna gain to that of the second nearest BS, as well as the PDF of the ratio of the corresponding path-losses are derived as insightful intermediate results. Subsequently, the corresponding analytical expressions for the cumulative distribution function (CDF) of the achieved signal-to-interference ratio (SIR) are obtained in exact form.

The remainder of the paper is organized as follows. In Section II, the angular distance distributions in a 2D finite HPPP are derived as preliminaries. In Section III, the system model is analytically presented and both the association policies and the beam selection schemes for the UE are proposed. In Section IV, the coverage probability analysis under the three association schemes, is presented while in Section V, the coverage probability analysis is conducted under the dominant interferer assumption. Section VI presents both the analytical and the numerical results, while Section VII concludes the paper.

 \section{Mathematical Preliminary: Angular Distances }
 In this section, the mathematical constructs of angular distances in 2D HPPP wireless networks  defined over a finite region, are presented. These will provide a firm foundation for the understanding and analysis of the coverage probability in mmWave cellular networks.
 
\subsection{Angular Distance Distributions}
 As stated in \cite{kanatas}, the investigation of the distributions of the angular distances is based on defining the $n$th nearest point in angle $\phi$ from a reference line. Consider a homogeneous PPP $\Phi$ with intensity $\lambda$  over $\mathbf{b}(o,r)$, where $ \mathbf{b}(o,r)$ denotes a ball of a finite radius $r$ centered at the origin $o$. Let $\phi_n$ denote the random variable representing the angular distance from an arbitrary reference line to the $n$th nearest point of $\Phi$ in the coordinate $\phi$, as shown in Fig. 1a. Without loss of generality, the reference line is assumed to be the $x$-axis.  Let $\mathcal{A}_{r,n}$ denote the event that $\mathbf{b}(o,r)$ contains at least $n$ points of $\Phi$.
 
 \textbf{Lemma 1.} \emph{Conditioned on $\mathcal{A}_{r,n}$, the PDF of the $n$th nearest point in angular distance  in $\mathbf{b}(o,r)$}, is given by
 \begin{equation} \label{eq1}
  f_{\phi_n | \mathcal{A}_{r,n}}(\phi) = \frac{(\lambda r^2)^n  \phi^{n-1}}{2^n  (\Gamma(n)-\Gamma(n,\lambda \pi r^2))} e^{-\frac{\lambda \phi r^2}{2}},  \quad \phi \in [0, 2\pi], 
\end{equation}
\emph{where $\Gamma(\cdot)$ is the Gamma function defined as in~\cite[eq. (8.310.1)]{Ryzhik}  and  $\Gamma(\cdot,\cdot)$ denotes the incomplete gamma function defined as in~\cite[eq. (8.350.2)]{Ryzhik}.} 
 
 \textit{Proof.} Let $W(\phi,r)$ denote a disk sector with dihedral angle $\phi$ and radius $r$. The area of the disk sector is $|W(\phi,r)|= \frac{\phi r^2}{2}$.  The conditional complementary CDF (CCDF) of $\phi_n$ is given by 
  \begin{equation} 
  \begin{split}
  \Tilde{F}_{\phi_n | \mathcal{A}_{r,n}}(\phi) & = \mathbb{P}[\Phi(W)<n | \Phi(\mathbf{b}(o,r)) \geq n] \\
  & = \frac{\mathbb{P}[\Phi(W)<n , \Phi(\mathbf{b}(o,r)) \geq n]}{\mathbb{P}[\Phi(\mathbf{b}(o,r)) \geq n]} \\
  & = \frac{\sum_{k=0}^{n - 1} P_k(\phi,r) \Big(1-\sum_{l=0}^{n -k- 1} Q_l(\phi,r)\Big)}{\sum_{k=n}^{\infty} P_k(2 \pi,r)}, 
  \end{split}
\end{equation}
 where $P_k(\phi,r) = e^{-\frac{\lambda \phi r^2}{2}} \frac{(\frac{\lambda \phi r^2}{2})^k}{k!}$,  $Q_l(\phi,r) = e^{-\lambda( \pi r^2 - \frac{\phi r^2}{2})} \frac{(\lambda r^2(\pi-\frac{\phi}{2}))^l}{l!}$ and $\Phi(\cdot)$ is a counting measure denoting the number of points of the point process $\Phi$ falling in an area. Now, the PDF of the conditional distance distribution is given by $f_{\phi_n | \mathcal{A}_{r,n}}(\phi)=-\frac{ {\rm{d}} \Tilde{F}_{\phi_n | \mathcal{A}_{r,n}}(\phi)}{{\rm{d}} \phi}$. Along similar conceptual lines as the proof presented in \cite[Corollary 2.3]{FHPPP}, $f_{\phi_n | \mathcal{A}_{r,n}}(\phi)$ can be obtained as 
\begin{equation} 
  f_{\phi_n | \mathcal{A}_{r,n}}(\phi) = \frac{ \frac{\lambda  r^2}{2} P_{n-1}(\phi,r) \sum_{k=0}^{\infty} Q_k(\phi,r)}{\sum_{k=n}^{\infty} P_k(2 \pi,r)}. 
\end{equation}
 By noticing that $\sum_{k=0}^{\infty} \frac{x^k}{k!} \myeqdef e^{x}$ , $\sum\limits_{\substack{k=n \\ n>0}}^\infty  \frac{x^k}{k!} \myeqdef \frac{e^{x} (\Gamma(n)-\Gamma(n,x))}{\Gamma(n)}$ and after applying some manipulations, we obtain (1).  \QEDA
 
 \begin{figure}[!t]
  \begin{subfigure}[b]{0.35\columnwidth}
    \includegraphics[trim=0 0 0 0,clip,width=\textwidth]{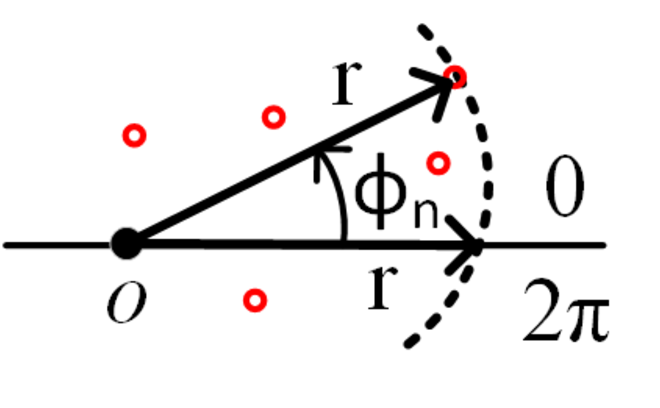}
    \caption{$\phi_n \in [0,2 \pi]$}
    \label{fig:1h}
  \end{subfigure}
  \hfill %%
  \begin{subfigure}[b]{0.35\columnwidth}
    \includegraphics[trim=0 0 0 0,clip,width=\textwidth]{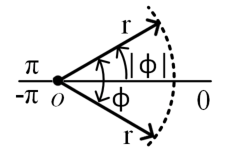}
    \caption{$|\phi| \in [0, \pi]$}
    \label{fig:2s}
  \end{subfigure}
  \caption{The $n$th nearest point in angular distance in a disk sector.}
\end{figure}

In mmWave cellular networks, it is often of interest the absolute angular distance, $|\phi|$, for the coordinate $\phi$. This definition is particularly useful when dealing with angular distances from the direction of the maximum directivity in an antenna beampattern \cite{kanatas}. Let $|\phi| \in [0, \pi]$, as shown in Fig. 1b. 

 \textbf{Corollary 1.}  \emph{Conditioned on $\mathcal{A}_{r,n}$, the PDF  $f_{|\phi_n| |  | \mathcal{A}_{r,n}}(|\phi|)$} is given by

\begin{equation} \label{eq4}
  f_{|\phi_n| | \mathcal{A}_{r,n}}(\phi) = \frac{(\lambda r^2)^n |\phi|^{n-1}}{ \Gamma(n)-\Gamma(n,\lambda \pi r^2)} e^{-\lambda |\phi| r^2}, \quad  |\phi| \in [0, \pi].
\end{equation}

\textit{Proof.} Letting  $\phi = 2 |\phi| \in [0, 2\pi]$, then $W(|\phi|,r) = |\phi| r^2$. Then, the proof follows the same steps as the proof for obtaining (\ref{eq1}). \QEDA 

\subsection{ Joint Distributions }
As mentioned in Section I.A, a UE may now be connected to the closest BS in angular distance. In this case, the dominant interfering BS may not necessarily be the closest one to the UE, but the closest BS in angular distance w.r.t the line of communication link. This inspires the investigation of joint distributions of angular distances. 

 Let $|\varphi_n|$ denote the absolute angular distance from $o$ to the $n$th nearest point in absolute angular distance. 
 
\textbf{Lemma 2.}  \emph{Conditioned on $\mathcal{A}_{r,2}$, the joint PDF of $|\varphi_1|, |\varphi_2|$ is given by}
\begin{equation} \label{eq5}
  f_{|\varphi_1|, |\varphi_2| \big| \mathcal{A}_{r,2} }(|\varphi_1|,|\varphi_2|) = \frac{(\lambda  r^2)^2\,  e^{-\lambda  r^2 ( |\varphi_2|-\pi)}}{e^{-\lambda \pi  r^2}- \lambda \pi  r^2 -1}, 
\end{equation}
 \emph{where  $|\varphi_1| \in [0,\pi]$ and  $|\varphi_2| \in [|\varphi_1|,\pi]$.}

\textit{Proof.} See Appendix A. \QEDA

\section{ System Model }
This section presents the considered mmWave cellular network and the key modeling assumptions.

\subsection{Network Model}
Consider a mmWave downlink cellular network, where the locations of the BSs is modeled as a HPPP $\Phi_{bs} \subset \mathbb{R}^2$ with intensity $\lambda_{bs}$. Let $(r_{x},\phi_{x})$ denote the location of a BS located at $x \in \Phi_{bs} $, in terms of polar coordinates. The locations of UEs are  independently distributed according to some stationary point process $\Phi_{ue}$. Also, a BS serves one UE at a time per resource block and all BSs are assumed to transmit at the same power $p$. Let $(r_{s},\phi_{s})$ denote the location of the serving BS $x_0 \in \Phi_{bs}$ at a given time. Without loss of generality, the receiving UE is assumed to be located at the origin $\mathbf{o} = (0,0)$ at that time.  After averaging the performance of this UE over $\Phi_{bs}$, the receiving UE becomes the typical receiver, which will be interchangeably referred to as just the {\em receiver} in this paper. 

\begin{figure}
    \centering
    \includegraphics[keepaspectratio,width=0.8\linewidth]{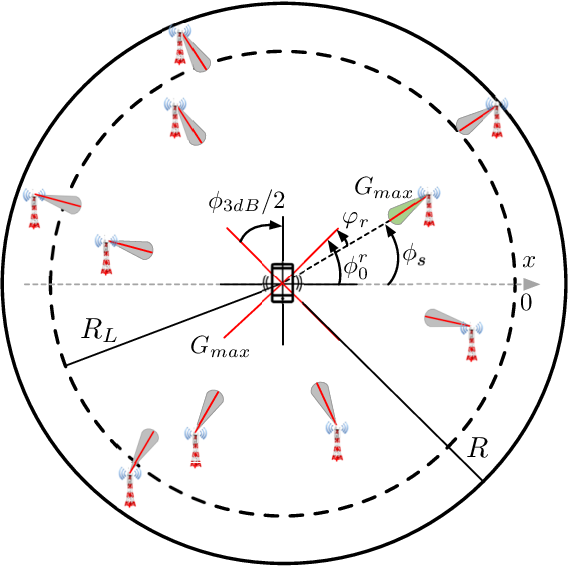}
     \caption{Illustration of the system model.}
    \label{fig:Fig1}
\end{figure}

 \subsection{Beamforming and Antenna Modeling}
 A BS located at $x \in \Phi_{bs}$ is assumed to exploit beamforming techniques to communicate with the receiver.  In this work, a worst-case interference scenario is considered in the mmWave network as the maximum gain of all BSs’ antennas is assumed to be always directed towards the receiver.  Please note that similar scenarios have already been employed in the literature \cite{WC2}. The receiver is assumed to be equipped with an antenna array able to produce $2^m$ receiving beams. The maximum gain directions of these beams, i.e., the centers of the corresponding 3 dB beamwidths, are given by $\phi_m=\frac{2 \pi}{2^m}$ with $m \in \mathbb{N}$. Please note that the maximums of the BS and UE beams will not necessarily be aligned because of the discretization on the UE side, which is the reason angular distances appear in the analysis. In this work, a realistic 3GPP-based antenna pattern recommended for 5G mmWave communications~\cite{pattern} is adopted for the receiver and therefore, the actual antenna gain within the 3dB beamwidth range is not constant. When directed towards $\phi_0^{r}$, this antenna has a radiation power pattern (in dB) given by
  \begin{equation}\label{eq6}
       G_{3gpp}(\phi-\phi_0^{r}) = G_{max}-{\rm{min}}\Bigg\{ 12\Big(\frac{\phi-\phi_0^{r}}{\phi_{3dB}} \Big)^2, SLA \Bigg\}, 
 \end{equation}
where $\phi_{3dB}$ is the 3dB beamwidth of the receiver's antenna and $SLA=30$dB is the front-to-back ratio. The multiplicative antenna gain factor is denoted by $g_{3gpp}(\phi-\phi_0^{r})$. In this work, the direction of maximum gain $\phi_0^{r}$ is modeled as a discrete random variable with probability mass function (pmf) given by 
\begin{equation}\label{eq7}
\phi_0^{r} = \left\{\begin{array}{ll} 
   \phi_0^{1} = \frac{\pi}{2^m},  & \quad   \mbox{w.p.} \, \, \, p_{\phi} = \frac{1}{2^m} \\
   \phi_0^{2} =  \frac{\pi}{2^m} +  \frac{\pi}{2^{m-1}},  &  \quad  \mbox{w.p.} \, \, \, p_{\phi} = \frac{1}{2^m} \\
   \vdots \\
   \phi_0^{2^m} =  \frac{\pi}{2^m} + \underbrace{\frac{\pi}{2^{m-1}}+...+\frac{\pi}{2^{m-1}}}_{(2^m - 1) \, {\rm{terms}}} ,  & \quad \mbox{w.p.} \, \, \, p_{\phi} = \frac{1}{2^m}
\end{array}\right.,
\end{equation}
A representative example of $\phi_0^{r}$ is shown in Fig. 2, with $2^m = 4$ beams enabled at the UE. Unfortunately,  the use of the array pattern of (\ref{eq6}) may lead to extremely intractable analysis \cite{Approx}. For this reason, the array pattern is approximated by the two-branch expression  given by \cite{Approx3gpp}, 
\begin{equation}\label{eq8}
g(\phi-\phi_0^{r}) = \left\{\begin{array}{ll} 
   g_{max} 10^{-\frac{3}{10} \Big(\frac{2 (\phi-\phi_0^{r})}{\phi_{3dB}}\Big)^2}, \,\,  |\phi-\phi_0^{r}| \leq \phi_A \\
   g_s,  \quad \quad \quad \quad \quad \quad   \phi_A \leq |\phi-\phi_0^{r}| \leq \pi
\end{array}\right.
\end{equation}
where $g_{max}$ and $g_s$ are the multiplicative gain factors of $G_{max}$, and $G_s= G_{max}-SLA$, correspondingly, whereas $\phi_A=(\phi_{3dB}/2) \sqrt{(10/3) \log_{10}(g_{max}/g_s)}$.

 \subsection{Blockage Model}
 As seen from the receiver location, a BS can be either LOS or non-LOS (NLOS). While mmWave signals are significantly affected by blockages~\cite{Heath}, proper modeling is essential to capture the effect of blockages on network performance. In this work, an LOS ball model is adopted since it provides a better fit with real-world blockage scenarios~\cite{Andrews},  which considers NLOS interfering links.  Hence, the performance analysis for the receiver is restricted to $\mathbf{b}(\mathbf{o},R)$, where $\mathbf{b}(\mathbf{o},R)$ denotes a ball of radius $R$  centered at the origin $\mathbf{o}$ and thus, to an almost surely (a. s.) finite HPPP $\Psi_{bs} = \Phi_{bs} \cap \mathbf{b}(\mathbf{o},R)$. Given a BS located at $x \in \Psi_{bs}$, the propagation between the BS and the receiver is LOS if $ \|x\| < R_{L}$, where $R_{L}$ is the maximum distance for LOS propagation and $\|\cdot\|$ denotes the Euclidean norm.

 % The received signal power from the BSs outside the LOS ball is considered negligible due to the severe path loss imposed by the blockages \cite{maxpower1}, \cite{NLOS}, and the authors in \cite{Heath2}, have shown that NLOS links have negligible effect on the system coverage performance in dense mmWave networks. 

 \subsection{Path Loss and Channel Models}
The LOS and NLOS channel conditions induced by the blockage effect are characterized by different path-loss exponents, denoted by $\alpha_L$ and $\alpha_N$, respectively. Similar to~\cite{Baccelli}, typical values of these path-loss exponents
are $\alpha_L \in [1.8, 2.5]$ and $\alpha_N \in [2.5, 4.7]$. Then, following the standard power-law path-loss model for the path between the receiver and a BS located at $x \in \Psi_{bs}$, the random path loss function  is given by
\begin{equation}
l(\|x\|)= \left\{
\begin{array}{ll}
       K \|x\|^{-\alpha_L}, & \|x\| < R_L \\
       K \|x\|^{-\alpha_N}, & R_L \leq \|x\| \leq R ,\\
\end{array} 
\right.    
\end{equation}

where $K = \Big(\frac{c}{4 \pi f_c}\Big)^2$ with $c$  being the speed of light and $f_c$ the carrier frequency.

The channels between a BS located at $x \in \Psi_{bs}$ and the receiver undergo Nakagami-\textit{$m_u$} fading, which is a generalized model for representing a wide range of fading environments. The parameter $m_u$ is restricted to integer values for analytical tractability. The channel fading gain $h_u$ is the fading power for the channel in LOS condition. The shape and scale parameters of $h_u$ are $m_u$ and $1/m_u$, respectively, i.e., $h_u \sim$ Gamma $\big(m_u, \frac{1}{m_u}\big)$. Here, $u \in \{s,x\}$, where $s$ denotes the link between the receiver and the serving BS, and $x$ stands for the link between the receiver and the interfering BSs at $x \in \Psi_{bs} \setminus \{x_0\}$. The PDF of $h_u$ is given by
\begin{equation} \label{eq9}
  f_{h_u}(w) = \frac{m_u^{m_u} w^{m_u-1}}{\Gamma(m_u)}\exp{(-m_u w)}. 
\end{equation}
Note that $\mathbb{E}[h_u]=1$, where $\mathbb{E}[\cdot]$ denotes expectation. Also, the values of $m_s$ are restricted to integers for analytical tractability.

\subsection{User Association and Beam Selection Policy}
The association policy used in real networks is often based on the maximum received power. The UE is agnostic to the conditions that provide the maximum power to the receiver. One of the main contributions of this paper is the adoption of an association policy, namely \textit{Policy 1}, that accurately captures the real conditions. This is achieved by jointly considering both polar coordinates in the calculation of received power. With the introduction of the angular coordinate,  this work investigates the use of a similar to the minimum Euclidean distance-based criterion, namely, the minimum angular distance-based one. This is henceforth termed as {\em Policy 2}. Nevertheless, for system performance analysis purposes it is highly desirable and widely accepted to consider simplified policies. Therefore, a commonly utilized policy, namely \textit{Policy 3}, is the one that assumes only the Euclidean distance criterion. However, this policy ignores the effect of angle-dependent antenna gains and inherently imposes a discrepancy from real received power in the calculations. Consequently, Policies 2 and 3 serve as baseline schemes for comparison to Policy 1 which corresponds to  the realistic maximum power association scheme.  We are particularly interested in understanding in which regimes can Policy 2 or 3 closely describe the performance of Policy 1, which is clearly more realistic but analytically more complicated compared to the other two.  Moreover, because of distance-based path-loss, the candidate serving BSs of the receiver are assumed to lie within $\mathbf{b}(\mathbf{o},R_L)$. From amongst these BSs, the serving BS is determined using the beam management procedure of 5G NR as discussed above. Therefore, the serving BS can lie anywhere within $\mathbf{b}(\mathbf{o},R_L)$. Finally, all the association decisions are assumed to take place within the coherence time of the channel. 

\subsubsection{Policy 1: Maximum Power-based Association Scheme}
In this scheme, the receiver is associated with the BS providing the largest average received power by jointly considering both the Euclidean and the angular distance of the BSs. Specifically, during this association procedure, the best receiving beam is selected by the receiver for reception. According to the beam management procedure in 5G NR, this task is the so-called beam selection procedure. In particular, all BSs periodically transmit the beamformed reference signals, either channel state information reference signals (CSI-RS) or synchronization signal blocks (SSBs), that may cover the entire set of available directions according to the receivers' needs. The measurements in mmWave networks related to the initial access are based on the SSBs. The receiver in the proposed framework monitors the reference signals and forms a list of \textit{candidate serving BSs}, being those BSs with the largest SNR and exceed a predefined threshold, for each beam. Subsequently, the \textit{serving BS}, defined as the BS providing the largest SNR among \textit{all} candidate serving BSs, is chosen for transmission. The corresponding receiver's beam in which the serving BS was identified, is selected as the \textit{receiving beam}.

 A key assumption in this policy is that the BS has perfect CSI knowledge of the uplink and thus, a perfect alignment of the BS beam maximum gain direction with the line toward the receiver is achievable. However, the direction of the maximum gain of the receiver's antenna is not fully aligned with the direction of maximum gain of each BS transmitting beam. Let $\varphi_i^x$ denote the  angular distance between the direction of the line connecting the receiver and a BS $x \in \Psi_{bs}$ and the direction, $\phi_0^i$, of maximum directivity of a receiver's beam $i$. Therefore, for a BS $x \in \Psi_{bs}$, $\phi_x$ denotes its polar coordinate, and $\varphi_i^x$ is given by $\varphi_i^x=|\phi_0^{i}-\phi_x|$. By further considering the Euclidean distances of the BSs from the origin, $r_x$, the location of the serving BS and the receiving beam are chosen by the receiver as
\begin{equation}  \label{eq11} 
  (x_0,i) = \underset{\begin{subarray}{c}
  x \in (\Psi_{bs} \cap \mathbf{b}(\mathbf{o},R_L)) \\
  i = 1:2^m
  \end{subarray}}{\operatorname{argmax}}\Big\{g(\varphi_i^x)r_x^{-\alpha_L}\Big\}.
\end{equation}

Having selected $x_0$ and $i$, the angle $\varphi_i^x$ is denoted as $\varphi_r$ in Fig. 2. In this case, $\varphi_{r}$ can be modeled as a uniform random variable, i.e., $\varphi_{r} \sim U[0,\frac{\phi_{3dB}}{2}]$, where $\phi_{3dB}$ denotes the half-power beamwidth of the user receiving beam.
Clearly, the serving BS may not necessarily be the nearest one to the receiver. Instead, it may lie close to the direction of the maximum directivity gain of a receiver's beam and thus providing maximum received power. Eq. (11) indicates that the receiver performs scanning for each beam to identify the serving BS by jointly considering both the angular distance of the BSs, from the direction of each beam's maxima, and the Euclidean distance of the BSs, respectively to maximize the received power. 

\subsubsection{Policy 2: Minimum Angular Distance Association Scheme}
In this scheme, the receiver is attached to the closest BS in angular distance  within $\mathbf{b}(\mathbf{o},R_L)$. Specifically, for each beam, the receiver performs exhaustive scanning and calculates the minimum angle between the direction of the beam's maxima and the direction of the line connecting the receiver and each BS. Then, from a list of $2^m$ angles, the receiver attaches to the BS with the minimum angular distance $\varphi_c$ and the corresponding beam is selected as the receiving beam. Mathematically, the location $x_0$ of the serving BS and the beam $i$ are selected by the receiver as 
\begin{equation} \label{eq13} 
 (x_0,i) = \underset{\begin{subarray}{c}
 x \in (\Psi_{bs} \cap \mathbf{b}(\mathbf{o},R_L)) \\
  i = 1:2^m
  \end{subarray}}{ \operatorname{argmin}}  \Big\{   \{|\phi_0^{i}-\phi_x|\}\Big\},   
\end{equation}
and $\varphi_c$ is given by 
\begin{equation} \label{eq14} 
 \varphi_c =  \underset{i = 1:2^m} {\operatorname{min}}  \Big\{ \underset{x \in (\Psi_{bs} \cap \mathbf{b}(\mathbf{o},R_L))}{  \operatorname{min}} \{|\phi_0^{i}-\phi_x|\}\Big\}. 
\end{equation}
 
\subsubsection{Policy 3: Minimum Euclidean Distance Association Scheme}
In this scheme, the receiver is associated with the nearest BS  in Euclidean distance within $\mathbf{b}(\mathbf{o},R_L)$. In this case, by adopting conventional beam management-based beam steering techniques, the maximum gain of the receiver’s antenna is assumed to be directed towards the serving BS and therefore perfect beam alignment is assumed. Without loss of generality, due to the isotropy and the stationarity  of the $\Psi_{bs}$, the reference line can be considered to be along the $x$-axis and passing through the origin. In this case, the location $x_0$ of the serving BS is chosen by the receiver as $x_0 = \underset{ x \in (\Psi_{bs} \cap \mathbf{b}(\mathbf{o},R_L))} {\operatorname{argmin}}\{r_x\}$.

%During an SSB burst, the beam pointing towards the serving BS is chosen by the receiver as the receiving beam. Consequently, during the association procedure, both the BSs and the receiver communicate via their main lobes i.e., the 3dB beamwidth part of the beam and not via side lobes. In other words, the direction of the 3dB beamwidth part of a beam will always be preferred for communication instead of the direction of the side lobe of another beam.  

\subsection{Signal-to-interference-plus-noise Ratio }
The SINR under the three association schemes is now defined. Under Policy 1, the received SINR is given by 
\begin{equation} \label{eq15} 
{\rm SINR} = \frac{p \, h_s \, g_{max} \,  g(\varphi_r) \, l(\|x_0\|)}{ I  + \sigma^2}, 
\end{equation}
where $p$ is the transmitted power and $I$ refers to the aggregate interference power and is given by $I = \sum_{x \in \Psi_{bs}^{!}} p \, h_x \, g_{max} \,  g_{3gpp}(\varphi_I) \, l(\|x\|)$, $ \Psi_{bs}^{!} = \big\{ \Psi_{bs} \setminus \{x_0\}  \big\}$,  $\varphi_I$ is defined  as $\varphi_I =|\phi_0^{r}-\phi_x|$ and $\sigma^2$ is the additive white Gaussian noise power. Please note that, $\phi_0^{r}$ also determines the direction of the reference line in this policy.

Under Policy 2, the received SINR is given by 
\begin{equation} \label{eq16} 
{\rm SINR} = \frac{p \, h_s \, g_{max} \,  g(\varphi_c) \, l(\|x_0\|)}{ I  + \sigma^2}, 
\end{equation}
where $I$  is given by $I = \sum_{x \in \Psi_{bs}^{!}} p \, h_x \, g_{max} \,  g_{3gpp}(\varphi_I) \, l(\|x\|)$  and $\varphi_I$ is defined  as $\varphi_I = |\varphi_c-\phi_x|$. Please note that, $\varphi_c$ also determines the direction of the communication link in this policy.  

Under Policy 3, the received SINR is given by 
\begin{equation} \label{eq17} 
{\rm SINR} = \frac{p \, h_s \, g_{max} \,  g_{max} \, l(\|x_0\|)}{ I  + \sigma^2}, 
\end{equation}
where $I$  is given by $I = \sum_{x \in \Psi_{bs}^{!}} p \, h_x \, g_{max} \,  g_{3gpp}(\phi_x) \, l(\|x\|)$.

\section{Coverage Probability Analysis}
In this section, performance analysis in terms of coverage probability, $\mathcal{P}_{c}(\gamma) \triangleq  \mathbb{P}({\rm SINR} > \gamma)$, is conducted under the three association schemes. 

\subsection{Coverage Probability Under Maximum Power Association Scheme}
To obtain $\mathcal{P}_{c}(\gamma)$ in this scheme, the PDF of the maximum received power must first be derived.  Let $S_{x}^{i} $ denote the received power from a BS located at $x \in (\Psi_{bs} \cap \mathbf{b}(\mathbf{o},R_L))$ measured w.r.t to the $i$-th beam. Then, $S_{x}^{i} $ is given by 
\begin{equation} 
S_{x}^{i} = g(|\phi_0^{i}-\phi_x|)\|x\|^{-\alpha_L} =  g(\underbrace{|\phi_0^{i}-\phi_x|}_{\varphi_i^x}) r_x^{-\alpha_L} 
\end{equation} 
and the maximum received power $S$ is obtained as   
\begin{equation}       
\begin{split}
S &= \underset{\begin{subarray}{c}
  x \in (\Psi_{bs} \cap \mathbf{b}(\mathbf{o},R_L)) \\
  i = 1:2^m
  \end{subarray}}{\operatorname{max}}\{S_{x}^i\} \\ 
 & \myeqa \underset{x \in (\Psi_{bs} \cap \mathbf{b}(\mathbf{o},R_L))} {\operatorname{max}}\{\underbrace{g(\varphi^x)r_x^{-\alpha_L}}_{S_{x}}\} 
  \myeqb g(\varphi_r)r_s^{-\alpha_L},  
\end{split}
\end{equation}  
where (a) follows  after maximization of $\{g(\varphi_i^x)\}_{_{x \in (\Psi_{bs} \cap \mathbf{b}(\mathbf{o},R_L))}}^{i=1:2^m}$ over $i$, i.e., for each $x$, the best beam is determined  and (b) follows from maximization over $x$ with $r_s = \|x_0\|$. Due to exhaustive scanning of the receiver in the whole $\mathbf{b}(\mathbf{o},R_L)$  and conditioned on $\mathcal{A}_{R_L,1}$ in order for $S$ to be meaningfully defined, the serving BS may lie anywhere in $\mathbf{b}(\mathbf{o},R_L)$\footnote{ Conditioned on $\mathcal{A}_{R_L,1}$, it is ensured that at least one LOS BS will always exist. Otherwise, the receiver would try to connect to the strongest NLOS BS, which is not of much practical interest because of significantly weaker NLOS links and is hence beyond the scope of the paper.} Therefore,  $r_x$ are independent and identically distributed (i.i.d.) in $\mathbf{b}(\mathbf{o},R_L)$ with PDF $f_{r_x}(r)$ of each element given by $f_{r_x}(r) = \frac{2 r}{R_L^2}$, $r \in [0,R_L]$.

\textbf{Lemma 3.} \emph{ Conditioned on $\mathcal{A}_{R_L,1}$, the PDF of the maximum received power $S$ is given by }
\begin{equation}  
f_{S| \mathcal{A}_{R_L,1}}(s_0) =  \frac{ \lambda_{bs} \pi R_L^2 \,f_{S_{x}}(s_0) \,e^{\lambda_{bs} \pi R_L^2 \big( F_{S_{x}}(s_0) - 1 \big) }}{1 - e^{- \lambda \pi R_L^2}},  
\end{equation}  
\emph{ with $s_0 \in [w_{min}, \infty)$, and $f_{S_{x}}(w)$ is given by }
\begin{equation}  
 f_{S_{x}}(w)= \int_{g_{3dB}}^{\psi(w)}\frac{1}{ x}f_{g(\varphi^x)}(x) f_{r_x^{-\alpha_L}}\Big(\frac{w}{x}\Big) {\rm d} x,  
\end{equation}  
\emph{with $w \in [w_{min}, \infty)$, and $f_{r_x^{-\alpha_L}}(x) = \frac{2 (\frac{1}{x})^{\frac{\alpha_L +2}{\alpha_L}}}{\alpha_L R_L^2}$,  $f_{g(\varphi^x)}(g) = \frac{1}{ln(10) x}\frac{10}{12 \sqrt{\frac{G_{max}-10 log(x)}{12}}}$, $g_{3dB} = 10^{\frac{G_{max}-3}{10}}$, $w_{min}=g_{3dB}R_L^{-\alpha_L}$ and $\psi(w) = {\rm{min}}\Big\{\frac{w}{R_L^{-\alpha_L}}, g_{max}\Big\}$.}  

\textit{Proof.} See Appendix B. \QEDA

\begin{figure*}[!b]
\hrulefill
\begin{equation}
\begin{split}
  \mathcal{L}_{I}(s|S_{th}, \phi_0^r) &= {\rm{exp}}\Big(- \lambda_{bs} \iint_{\mathbf{\Omega}} \Big(1- \Big(1+\frac{s\, p\, K\, g_{max}\, g_{3gpp}(|\phi_0^r-\phi_x|)\, r_x^{-\alpha_L}}{m_x} \Big)^{-m_x}\Big)r_x  {\rm d} \phi_x {\rm d} r_x  \Big)\\
 & \times {\rm{exp}}\Big(- \lambda_{bs} \iint_{\mathbf{\Omega^{'}}} \Big(1- \Big(1+\frac{s\, p\, K\, g_{max}\, g_{3gpp}(|\phi_0^r-\phi_x|)\, r_x^{-\alpha_N}}{m_x} \Big)^{-m_x}\Big)r_x  {\rm d} \phi_x {\rm d} r_x  \Big),
\end{split}
\end{equation}
\end{figure*}

 \textbf{Remark 1.} \emph{Lemma 3 captures the misalignment error as a random variable within $g(\varphi_r)$ and thus, it is clearly different from deriving the conventional PDF of the minimum distance-based path-loss.}

\textbf{Lemma 4.} \emph{Conditioned on the maximum received power $S=S_{th}$  from the serving BS w.r.t. a given receiving beam, the conditional Laplace transform $\mathcal{L}_{I}(s|S_{th}, \phi_0^r)$ of the aggregate interference power distribution is given by (21) (shown at the bottom of the page), where $\mathbf{\Omega} = \{ (r_x,\phi_x) \in \mathbb{R}^2 | r_x^{min} \leq r_x \leq R_L, \, 0 \leq \phi_x \leq 2 \pi\}$, $\mathbf{\Omega^{'}} = \{ (r_x,\phi_x) \in \mathbb{R}^2 | r_x^{max}  \leq r_x \leq  R, \, 0 \leq \phi_x \leq 2 \pi\}$ and $r_x^{min} = {\rm{min}}\Big\{\Big(\frac{g_{3gpp}(|\phi_0^r-\phi_x|)}{S_{th}}\Big)^{\frac{1}{\alpha_L}} ,R_L\Big\}$ , $r_x^{max} = {\rm{max}}\Big\{\Big(\frac{g_{3gpp}(|\phi_0^r-\phi_x|)}{S_{th}}\Big)^{\frac{1}{\alpha_N}} ,R_L\Big\}$.}

\begin{figure*}
  \begin{subfigure}[t]{.5\linewidth}
    \includegraphics[trim=20 0 15 0,clip,width=\linewidth]{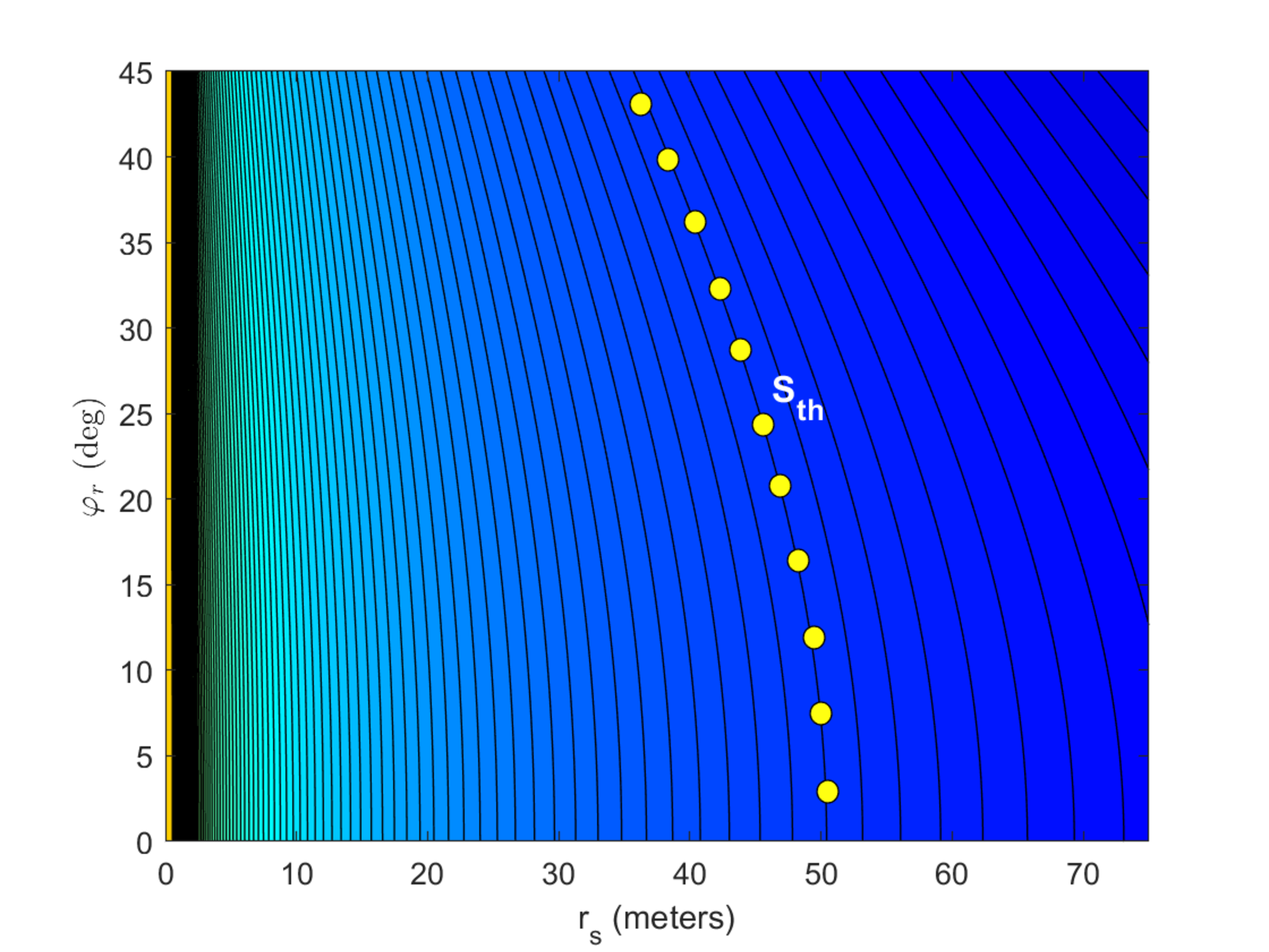}%
    \caption{Powersurface of maximum received power $S$, versus $r_s$ and $\varphi_r$.}%
\label{a}
  \end{subfigure}\hfil
  \begin{subfigure}[t]{.5\linewidth}
    \includegraphics[trim=20 0 20 0,clip,width=\linewidth]{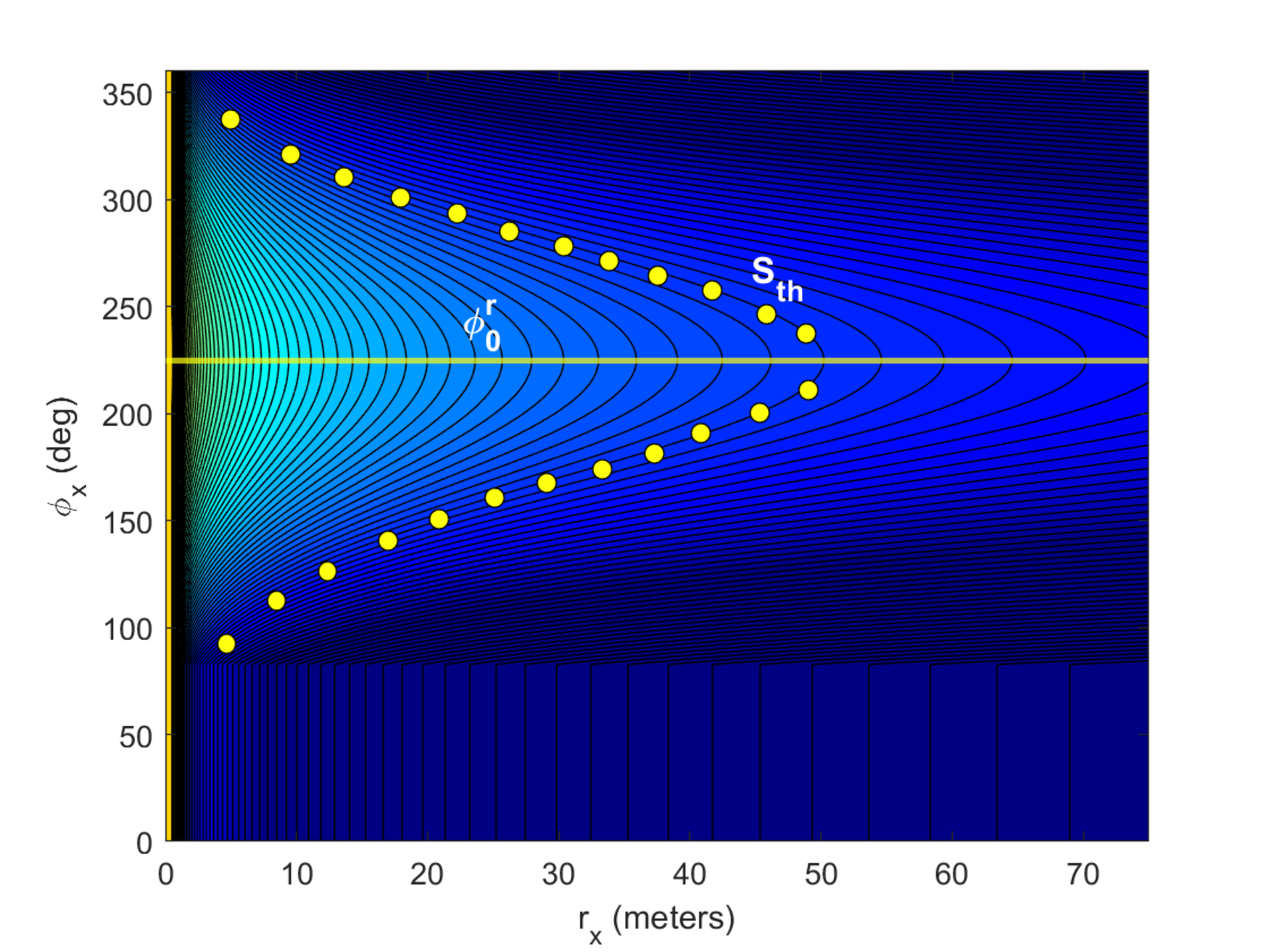}%
    \caption{Powersurface of received power $S_x= g(|\phi_0^r-\phi_x|)r_x^{-\alpha_L}$, versus $r_x$ and $\phi_x$.}%
  \end{subfigure}
   \caption{ Representative example of powersurfaces for the serving and interfering BSs corresponding to a receiver equipped with a 4-sectored antenna, $\phi_0^r= \pi/4+\pi/2+\pi/2$ and $S=S_{th}$. }
\end{figure*}

\textit{Proof.} Conditioned on $S = g(\varphi_r)r_s^{-\alpha_L}$, the location $x_0$ of the serving BS and the maxima $\phi_0^r$ of the receiving beam are known.  Once the serving BS has been determined, by appealing to Slivnyak’s Theorem, $\Psi_{bs}^{!}$ is a HPPP restricted to the set $\{\mathbf{\Omega} \cup \mathbf{\Omega^{'}}\}$. The set $\{\mathbf{\Omega} \cup \mathbf{\Omega^{'}}\}$ is defined by all possible locations of the interfering BSs, being those locations which lie outside the exclusion zone. Note that the exclusion zone is defined by all possible locations of $x_0$. However, the location $x_0$, given through $\varphi_r$, $r_s$, is a function of the maximum received power $S_{th}$. In Fig. 3a, a representative powersurface example of the maximum received power $S$ from $x_0$ in terms of $\varphi_r$ and $r_s$, is illustrated, by assuming a 4-sectored antenna, i.e, $\phi_{3dB}=\pi/2$. In Fig. 3b, the corresponding powersurface of the received power $S_x= g(|\phi_0^r-\phi_x|)r_x^{-\alpha_L}$ from $ x \in (\Psi_{bs} \cap \mathbf{b}(\mathbf{o},R_L))$ in terms of $r_x, \phi_x$, is shown, by assuming that the third beam has been chosen as the receiving beam and therefore $\phi_0^r= \pi/4+\pi/2+\pi/2$. Notice that for a given value of $S=S_{th}$ in Fig. 3a, the received power from the interfering BSs should be smaller than $S_{th}$ and the locations of $x \in \Psi_{bs}^{!}$ must lie outside the exclusion zone indicated by  the yellow markers. Also, notice that in Fig. 3b, $S_{th}$ indicates the maximum received power of an interfering BS. Considering the aforementioned, the minimum distance $r_x^{min}$  is given by 
\begin{equation}  \label{C1}
\begin{split}
&g_{3gpp}(|\phi_0^r-\phi_x|){r_x^{min}}^{-\alpha_L} < \underbrace{g(\varphi_r)r_s^{-\alpha_L}}_{S_{th}} \\
&\Leftrightarrow \frac{g_{3gpp}(|\phi_0^r-\phi_x|)}{S_{th}} < {r_x^{min}}^{-\alpha_L}\\
&\Leftrightarrow r_x^{min} > \Big( \frac{g_{3gpp}(|\phi_0^r-\phi_x|)}{S_{th}}\Big)^{\frac{1}{\alpha_L}}, 
\end{split}
\end{equation}
as  $\phi_x \in [0, 2\pi]$. Now, by observing from Fig. 3b that $r_x>R_L$ for some values of $S$,  ${r_x^{min}}$ is rewritten as $r_x^{min}={\rm{min}}\Big\{\Big( \frac{g_{3gpp}(|\phi_0^r-\phi_x|)}{S_{th}}\Big)^{\frac{1}{\alpha_L}},R_L\Big\}$.  Now, along similar conceptual lines, $r_x^{max}$ can be derived. Next, the Laplace transform can be obtained as described in Appendix C. \QEDA 

\textbf{Theorem 1.} \emph{ Conditioned on $\mathcal{A}_{R_L,1}$, the coverage probability of a receiver in a mmWave network inside $\mathbf{b}(\mathbf{o},R_L)$ under the maximum power-based association policy is given by}
\begin{equation} 
\begin{split}
&\mathcal{P}_{c}(\gamma) =\\
& \sum_{i=1}^{2^m} \sum_{k=0}^{m_s - 1} \int_{w_{min}}^{\infty} \frac{(-s)^k}{k!} { \Bigg[\frac{\partial^{k} \mathcal{L}_{I_{tot}}(s|S_{th}, \phi_0^i)}{\partial{s^k}}\Bigg]}_{s} f_{S| \mathcal{A}_{R_L,1}}(S_{th})  \\
& \times p_\phi {\rm d} S_{th},
\end{split}
\end{equation}
\textit{where} $s=\frac{m_s \gamma}{p\,g_{max}\,K\,S_{th}}$ \textit{and} $\mathcal{L}_{I_{tot}}(s|S_{th}, \phi_0^i) = \exp({-\sigma^2 s)}   \mathcal{L}_{I}(s|S_{th}, \phi_0^i)$.

%\begin{equation} 
%\begin{split}
%\mathcal{P}_{c}(\gamma) & =  \mathbb{P}\Big[ \frac{p \, h_s \, g_{max} \,  g(\varphi_r) \, l(\|x_0\|)}{ I  + \sigma^2} > \gamma \Big| r_s, \phi_s, \phi_0^r \Big]\\
%&=   \mathbb{P}\Big[ \frac{p \, h_s \, g_{max} \, K \,  S_{th}}{ I  + \sigma^2} > \gamma \Big| S_{th} , \phi_0^r \Big] \\
%& =  \mathbb{P}\Big[ h_s > \frac{\gamma (I + \sigma^2)}{p \, g_{max} \, K \, S_{th}} \Big| S_{th}, \phi_0^r\Big]\\
%&\myeqa  \sum_{k=0}^{m_s - 1} \frac{(-s)^k}{k!} { \Bigg[\frac{\partial^{k} \mathcal{L}_{I_{tot}}(s|S_{th}, \phi_0^r)}{\partial{s^k}}\Bigg]}_{s} \\
%& \myeqb \sum_{i=1}^{2^m} \sum_{k=0}^{m_s - 1} \int_{w_{min}}^{\infty} \frac{(-s)^k}{k!} { \Bigg[\frac{\partial^{k} \mathcal{L}_{I_{tot}}(s|S_{th}, \phi_0^i)}{\partial{s^k}}\Bigg]}_{s}\\
%& \times f_{S}(S_{th}) p_\phi {\rm d} S_{th}, 
%\end{split}
%\end{equation}

\textit{Proof.} The conditional coverage probability is first given by (24) (shown at the bottom of the next page) where (a) follows from the CCDF of $h_s$, the definition of incomplete gamma function for integer  values of $m_s$ and by using $ \mathbb{E}_{I_{tot}}[{\rm{exp}}(-s I_{tot}) {(s I_{tot})}^k]=(-s)^k\frac{\partial^{k} \mathcal{L}_{I_{tot}}(s)}{\partial s^k}$, and (b) follows from deconditioning over the maximum power $S$ and all possible receiving beam's maxima $\phi_0^r$, with the PDF $f_{S}(s_0)$ and the pmf $p_{\phi}$, respectively. \QEDA  

\begin{figure*}[!b]
\hrulefill
\begin{equation}
\begin{split}
\mathcal{P}_{c}(\gamma) & =  \mathbb{P}\Big[ \frac{p \, h_s \, g_{max} \,  g(\varphi_r) \, l(\|x_0\|)}{ I  + \sigma^2} > \gamma \Big| r_s, \phi_s, \phi_0^r \Big]=   \mathbb{P}\Big[ \frac{p \, h_s \, g_{max} \, K \,  S_{th}}{ I  + \sigma^2} > \gamma \Big| S_{th} , \phi_0^r \Big]   \mathbb{P}\Big[ h_s > \frac{\gamma (I + \sigma^2)}{p \, g_{max} \, K \, S_{th}} \Big| S_{th}, \phi_0^r\Big] \\
&\myeqa  \sum_{k=0}^{m_s - 1} \frac{(-s)^k}{k!} { \Bigg[\frac{\partial^{k} \mathcal{L}_{I_{tot}}(s|S_{th}, \phi_0^r)}{\partial{s^k}}\Bigg]}_{s}  \myeqb \sum_{i=1}^{2^m} \sum_{k=0}^{m_s - 1} \int_{w_{min}}^{\infty} \frac{(-s)^k}{k!} { \Bigg[\frac{\partial^{k} \mathcal{L}_{I_{tot}}(s|S_{th}, \phi_0^i)}{\partial{s^k}}\Bigg]}_{s} f_{S}(S_{th}) p_\phi {\rm d} S_{th}, 
\end{split}
\end{equation}
\end{figure*}

\subsection{Coverage Probability for minimum angular distance association scheme}  
In this scheme, the receiver searches and associates with the BS that has the minimum angular distance $\varphi_c$ among all minimum angular distances detected from each beam's maxima. In this case, the receiver may lie anywhere in $\mathbf{b}(\mathbf{o},R_L)$ and $\varphi_c$ also denotes the serving angular distance \footnote{Considering infinitely large LOS ball radius, the Euclidean distance between the receiver and the serving BS may tend to infinity. In general, this scheme underestimates the performance. In a similar vein, if the receiver is associated with the nearest BS in Euclidean distance, the serving BS may lie near the boundaries of the 3dB beamwidth of the receiver’s antenna, which is also not accurate and will underestimate the performance. The aforementioned conceptual gap is resolved by adopting Policy 1.}. Therefore, the serving distance $d_0 =\|x_0\|$ is now independent of the association policy and i.i.d. in $\mathbf{b}(\mathbf{o},R_L)$ with PDF  given by $f_{d_0}(d_0) = \frac{2 d_0}{R_L^2}$, $d_0 \in [0,R_L]$. The PDF of $\varphi_c$ is first derived.

\textbf{Lemma 5.} \emph{For a receiver equipped with $2^m$ sectors  and conditioned on $\mathcal{A}_{R_L,1}$ the PDF of the closest angular distance $\varphi_c$ is given by}
\begin{equation} 
\begin{split}
  f_{\varphi_c|\mathcal{A}_{R_L,1}}(\varphi_c) = \frac{\lambda_{bs} 2^m  R_L^2 {\rm{exp}}(-\lambda_{bs} 2^m \varphi_c R_L^2)}{1-e^{-\lambda_{bs} \pi R_L^2}},  
  \end{split}
\end{equation} 
\emph{for $ \varphi_c \in \big[0,\frac{\pi}{2^m}\big]$}.

\begin{figure*}[!b]
\hrulefill
\begin{equation}
\begin{split}
 \mathcal{L}_{I}(s|\varphi_c, \phi_0^r) &= {\rm{exp}}\Big(- \lambda_{bs}  \iint_{\mathbf{V}} \Big(1- \Big(1+\frac{s\, p\, K\, g_{max}\, g_{3gpp}(|\phi_0^r-\phi_x|)\, r_x^{-\alpha_L}}{m_x} \Big)^{-m_x}\Big) r_x  {\rm d} \phi_x {\rm d} r_x  \Big) \\
 &\times {\rm{exp}}\Big(- \lambda_{bs}  \iint_{\mathbf{V^{'}}} \Big(1- \Big(1+\frac{s\, p\, K\, g_{max}\, g_{3gpp}(|\phi_0^r-\phi_x|)\, r_x^{-\alpha_N}}{m_x} \Big)^{-m_x}\Big) r_x  {\rm d} \phi_x {\rm d} r_x  \Big),
\end{split}
\end{equation}
\end{figure*}

\textit{Proof.} Through manipulation of \eqref{eq4}, let $r=R_L$, $\lambda=\lambda_{bs}$, $\phi=2^{m+1} \varphi_c $,  $n=1$ and  $\varphi_c \in [0,\frac{\pi}{2^m}]$ so that $\phi \in [0, 2\pi]$. Then, $W(\phi,r) = \frac{\phi r^2}{2}= 2^m \varphi_c R_L^2 $. Next, the proof follows the same steps as the proof for deriving \eqref{eq4}.  \QEDA

%Given $\varphi_c$ and the maxima $\phi_0^r$, the conditional Laplace transform of the aggregate interference power distribution can be derived. 

\textbf{Lemma 6.} \emph{Conditioned on  $\varphi_c$ w.r.t.  $\phi_0^r$ , the conditional Laplace transform $\mathcal{L}_{I}(s|\varphi_c, \phi_0^r)$ of the aggregate interference power distribution is given by (26) (shown at the bottom of the next  page)}
\emph{where $\mathbf{V} = \{ (r_x,\phi_x) \in \mathbb{R}^2 | 0 \leq r_x \leq R_L, \, \varphi_c \leq \phi_x \leq 2 \pi\}$ and  $\mathbf{V^{'}} = \{ (r_x,\phi_x) \in \mathbb{R}^2 | R_L \leq r_x \leq R, \, \varphi_c \leq \phi_x \leq 2 \pi\}$}.

\textit{Proof.} As the distance of $x_0$ is independent of the association scheme, the locations of $x \in \Psi_{bs}^{!}$ can be anywhere in angular distance larger than the serving angular distance $\varphi_c$, and therefore let  $\mathbf{V} = \{ (r_x,\phi_x) \in \mathbb{R}^2 | 0 \leq r_x \leq R_L, \, \varphi_c \leq \phi_x \leq 2 \pi\}$. Then, the proof follows the same steps shown in Appendix C and hence it is omitted here. \QEDA 

%The exact expression of the coverage probability can now be derived in the following Theorem. 

\textbf{Theorem 2.} \emph{ Conditioned on $\mathcal{A}_{R_L,1}$,} the coverage probability of a receiver in a mmWave network inside $\mathbf{b}(\mathbf{o},R_L)$ under a minimum angular distance-based association policy is given by
 \begin{equation} 
\begin{split}
&\mathcal{P}_{c}(\gamma)= \sum_{i=1}^{2^m} \sum_{k=0}^{m_s - 1} \int_{0}^{\frac{\pi}{2^m}} \int_{0}^{R_L} \frac{(-s)^k}{k!} { \Bigg[\frac{\partial^{k} \mathcal{L}_{I_{tot}}(s|\varphi_c, \phi_0^i)}{\partial{s^k}}\Bigg]}_{s} \\
&\times p_\phi f_{d_0}(d_0) f_{\varphi_c|\mathcal{A}_{R_L,1}}(\varphi_c) {\rm d} d_0 {\rm d} \varphi_c,
\end{split}
\end{equation}
where $s=\frac{m_s \gamma}{p\,g_{max}\,K\,g(\varphi_c) d_0^{-\alpha_L}}$ and $\mathcal{L}_{I_{tot}}(s|\varphi_c, \phi_0^i) = \exp({-\sigma^2 s)}   \mathcal{L}_{I}(s|\varphi_c, \phi_0^i)$.

\textit{Proof.} Similar to the proof for deriving (23),  
\begin{equation} 
\begin{split}
&\mathcal{P}_{c}(\gamma)  =  \mathbb{P}\Big[ \frac{p \, h_s \, g_{max} \,  g(\varphi_c) \, l(\|x_0\|)}{ I  + \sigma^2} > \gamma \Big| \varphi_c, \phi_0^r, d_0 \Big]  \\
& \myeqa \sum_{i=1}^{2^m} \sum_{k=0}^{m_s - 1} \int_{0}^{\frac{\pi}{2^m}} \int_{0}^{R_L} \frac{(-s)^k}{k!} { \Bigg[\frac{\partial^{k} \mathcal{L}_{I_{tot}}(s|\varphi_c, \phi_0^i)}{\partial{s^k}}\Bigg]}_{s}\\
&\times p_\phi f_{d_0}(d_0) f_{\varphi_c|\mathcal{A}_{R_L,1}}(\varphi_c) {\rm d} d_0 {\rm d} \varphi_c, 
\end{split}
\end{equation}
where (a) follows from deconditioning over the conditioned random variables $\varphi_c, \phi_0^r, d_0$ of the conditional coverage probability. \QEDA 

\subsection{Coverage Probability for minimum Euclidean distance association scheme}  
In this scheme, the receiver associates with the closest BS in Euclidean distance. Once the communication link has been established, all $x \in \Psi_{bs}^{!}$ interfere to the receiver with AoAs $\phi_x \sim U[-\pi,\pi]$.

\begin{figure*}[!b]
\hrulefill
 \begin{equation} 
\begin{split}
 \mathcal{L}_{I}(s|r_1) &= {\rm{exp}}\Big(- \lambda_{bs} \iint_{\mathbf{Q}} \Big(1- \Big(1+\frac{s\, p\, K\, g_{max}\, g_{3gpp}(\phi_x)\, r_x^{-\alpha_L}}{m_x} \Big)^{-m_x}\Big) r_x  {\rm d} \phi_x {\rm d} r_x  \Big)\\
 & \times {\rm{exp}}\Big(- \lambda_{bs} \iint_{\mathbf{Q^{'}}} \Big(1- \Big(1+\frac{s\, p\, K\, g_{max}\, g_{3gpp}(\phi_x)\, r_x^{-\alpha_N}}{m_x} \Big)^{-m_x}\Big) r_x  {\rm d} \phi_x {\rm d} r_x  \Big),
\end{split}
\end{equation} 
\end{figure*}

\textbf{Lemma 7.} \emph{The Laplace transform of the aggregate interference power distribution conditioned on the serving distance $r_1$ is given by (29) (shown at the bottom of the page)}
\emph{where $\mathbf{Q} = \{ (r_x,\phi_x) \in \mathbb{R}^2 | r_1 \leq r_x \leq R_L, \, -\pi \leq \phi_x \leq  \pi\}$ and $\mathbf{Q^{'}} = \{ (r_x,\phi_x) \in \mathbb{R}^2 | R_L \leq r_x \leq R, \, -\pi \leq \phi_x \leq  \pi\}$}.

\textit{Proof.} The proof for deriving Lemma 7 follows similar steps as the one of Lemma 6. \QEDA

%The exact expression of the coverage probability for the receiver is now presented. 

\textbf{Theorem 3.} \emph{ Conditioned on $\mathcal{A}_{R_L,1}$, the coverage probability of a receiver in a mmWave network inside $\mathbf{b}(\mathbf{o},R_L)$ under minimum Euclidean distance association policy is given by}
 \begin{equation} 
\begin{split}
&\mathcal{P}_{c}(\gamma)  \\
& =  \sum_{k=0}^{m_s - 1} \int_{0}^{R_L} \frac{(-s)^k}{k!} { \Bigg[\frac{\partial^{k} \mathcal{L}_{I_{tot}}(s|r_1)}{\partial{s^k}}\Bigg]}_{s=\frac{m_s \gamma r_1^{\alpha_L}}{p\,K\,g_{max}^2}} \\ 
& \times f_{r_1|\mathcal{A}_{R_L,1}}(r_1)  {\rm d} r_1,
\end{split}
\end{equation}
\textit{where}  $\mathcal{L}_{I_{tot}}(s|r_1) = \exp({-\sigma^2 s)}   \mathcal{L}_{I}(s|r_1)$ and $f_{r_1| \mathcal{A}_{R_L,1}}(r_1) = \frac{2 \pi \lambda_{bs} r_1 e^{-\lambda_{bs} \pi r_1^2}}{1-e^{-\lambda \pi r^2}}$, $r_1 \in [0,R_L]$ .

\textit{Proof.} The proof for deriving Theorem 3 follows similar steps as the proof in Theorem 1. \QEDA

\section{Special Cases: Dominant Interferer Approach}
The dominant interferer approach has been widely used in the literature due to its usefulness when the exact analysis is too complicated or leads to unwieldy results. For instance, in \cite{dom1}$- \hspace{-0.15cm}$\cite{dom4}, the authors capture the effect of the dominant interferer while approximating the residual interference with a mean value. In this section, in order to understand the effect of the different potential definitions of the dominant interferer on the performance analysis, the coverage performance is investigated under the assumption of neglecting all but a single dominant interferer for  all policies. Note that in order to define the interferer as dominant, the latter is restricted to $\mathbf{b}(\mathbf{o},R_L)$. Accordingly, a performance comparison between the dominant interferer approaches with the exact performance of Policy 1, is conducted. To this end, the noise power is assumed to be negligible as compared to the aggregate interference experienced at the receiver, i.e., interference-limited scenarios \footnote{Please note, that since interference from other BSs is ignored, it results in stochastic dominance of the SIR as compared to the exact SIR of the respective policy, which implies that the dominant interferer approach yields a bound on the exact coverage probability of  Policy 1,} Policy 2 and Policy 3. are considered and coverage probability, i.e., $\mathcal{P}_{c}(\gamma) \triangleq  1 - F_{{\rm SIR}}(\gamma)$, analysis is conducted in terms of the achieved SIR.

 \subsection{Coverage Probability Under Policy 1}
By considering maximum power user association policy, the receiver associates with the BS providing the maximum power as described in subsection III.E. Then, clearly the dominant interferer is the BS which provides the second most powerful received power after the serving BS. The received SIR for this dominant interferer analysis of Policy 1 can be written as
\begin{equation}    
{\rm SIR} = \frac{ h_1 Pr_0}{g(|\varphi_2|) h_2 Pr_I},
\end{equation}
where $Pr_0=S$ and $Pr_I$ denotes the second most powerful received power. Once the receiver attaches
to the serving BS, $Pr_0$ and $Pr_I$ are no longer independent. In this case, the joint PDF of $Pr_0$ and $Pr_I$ is required for the coverage analysis, which is given next.  

\textbf{Lemma 8.} \emph{Conditioned on $\mathcal{A}_{R_L,2}$, the joint PDF of $Pr_0$ and $Pr_I$ is given by} 
\begin{equation}  
\begin{split}
&f_{Pr_0,Pr_I | \mathcal{A}_{R_L,2}}(s_0,s_I) \\
&= \frac{(\lambda_{bs} \pi R_L^2)^2 e^{\lambda_{bs} \pi R_L^2 \big(F_{S_{x}}(s_0)-1\big)} f_{S_{x}}(s_0) f_{S_{x}}(s_I)}{1-\Gamma(2, \lambda_{bs} \pi R_L^2)}, 
\end{split}
\end{equation}
\emph{where $s_0 \in [w_{min}, \infty]$ and  $s_I \in [w_{min}, s_0]$}. 

\textit{Proof.} By exploiting theory of Order Statistics \cite{orderstatistics}, and given that $k$ BSs ($k \geq 2$) exist in $\mathbf{b}(\mathbf{o},R_L)$, the joint PDF is first given by
\begin{equation}    
f_{Pr_0,Pr_I|k }(s_0,s_I) = k (k-1) (F_{S_{x}}(s_0)^{k-2} f_{S_{x}}(s_0) f_{S_{x}}(s_I). 
\end{equation}
Then, $f_{Pr_0,Pr_I | \mathcal{A}_{R_L,2}}(s_0,s_I)$ can be obtained along similar lines as the proof presented in Appendix A and thus it is omitted here. \QEDA

Having obtained $f_{Pr_0,Pr_I | \mathcal{A}_{R_L,2}}(s_0,s_I)$, it becomes mathematically convenient to statistically characterize the SIR as SIR = $ P \cdot Q$, where $P = \frac{Pr_0}{Pr_I}$ and $Q = \frac{h_1}{h_2}$. 

\textbf{Lemma 9.} \emph{Conditioned on $ \mathcal{A}_{R_L,2}$, the PDF of $P$ is given by} 
\begin{equation}    
f_{P|  \mathcal{A}_{R_L,2}}(p_0) = \int_{p_{min}(p_0)}^{\infty} p_0 f_{Pr_0,Pr_I |  \mathcal{A}_{R_L,2}}(p_0 s_I,s_I) {\rm{d}} s_I, 
\end{equation}
\emph{where $p_0 \in [1, \infty]$ and $p_{min}(p_0) = {\rm{max}}\Big\{\frac{w_{min}}{p_0}, w_{min} \Big\}$. } 

\textit{Proof.} The proof results directly from the formula for the PDF of ratio of two dependent random variables. \QEDA 

\textbf{Lemma 10.} \emph{Conditioned on $ \mathcal{A}_{R_L,2}$, the PDF of $Q$ is given by} 
\begin{equation}    
f_{Q| \mathcal{A}_{R_L,2}}(q) = \frac{q^{m_s-1}  m_s^{m_s} m_x^{m_x} \Gamma(m_s + m_x)}{\Gamma(m_s )\Gamma(m_x) (q m_s + m_x)^{(m_s + m_x)}} , 
\end{equation}
\emph{where $q \in [0, \infty]$}. 

\textit{Proof.} The proof results directly from the formula for the PDF of ratio of two independent random variables. \QEDA

\textbf{Proposition 1.} \emph{Conditioned on $ \mathcal{A}_{R_L,2}$, the coverage probability in the presence of a dominant interferer under Policy 1 is given by}
\begin{equation}    
\mathcal{P}_{c}(\gamma)  = 1 - \int_0^\gamma  \int_1^{\infty}  \frac{1}{x} f_{P|\mathcal{A}_{R_L,2}}(x) f_{Q|\mathcal{A}_{R_L,2}}\Big(\frac{x}{p_0}\Big) {\rm d} p_0 {\rm d} x.
\end{equation}

\textit{Proof.} The PDF of the SIR results directly from expressing the SIR given as a product of two random variables 
\begin{equation} 
f_{SIR} (x)  =  \int_1^{\infty}  \frac{1}{x} f_{P| \mathcal{A}_{R_L,2}}(x) f_{Q|| \mathcal{A}_{R_L,2}}(\frac{x}{p_0}) {\rm d} p_0.
\end{equation} 
Then, $F_{SIR}(\gamma) =  \int_0^\gamma f_{SIR} (x) {\rm d} x$. \QEDA

\subsection{Coverage Probability Under Policy 2}
By assuming that the receiver associates with the closest BS in angular distance, the dominant interferer is the second nearest BS in angular distance w.r.t the maxima of the receiving beam, as described in subsection III.E. Without loss of generality, the receiving beam's maxima is assumed to be along the $x$-axis.\footnote{As the scope of this section is merely to address the impact of the dominant interferer in the coverage performance of mmWave networks, the exhaustive scanning procedure for the association policy proposed in Section III.E is not considered here as the location of the dominant interferer is independent of the considered reference line.} In this case, the received SIR is given by 
\begin{equation}    
{\rm SIR} = \frac{g(|\varphi_1|) h_1 d_1^{-\alpha_L}}{g(|\varphi_2|) h_2 d_2^{-\alpha_L}},
\end{equation}
where $f_{d_i}(d_i) = \frac{2 d_i}{R_L^2}$, $i \in \{1,2\}$ denotes the distance of the serving BS and the dominant interferer, respectively, $|\varphi_1|, |\varphi_2|$ denote the nearest and the second nearest absolute angular distances of the serving and interfering BS from the direction of the receiving beam's maximum directivity, and $h_1=h_s, h_2=h_x$, for notational convenience, respectively. Note that in order to define the interferer as "dominant", the angle $|\varphi_2|$ is restricted to $|\varphi_2| < \phi_A$, i.e., the dominant interferer is assumed to fall in the mainlobe part of the receiver’s antenna pattern.   Let $T_{\phi_A,2}$ define the event that 2 BSs (assumed to lie in $\mathbf{b}(\mathbf{o},R_L)$) fall in the mainlobe part of the receiver's antenna. Note that $T_{\phi_A,2} \subseteq \mathcal{A}_{R_L,2}$.  Once the receiver attaches to the closest BS at $|\varphi_1|$, the angular distances $|\varphi_1|, |\varphi_2|$ are no longer independent. In this case, it is mathematically convenient to statistically characterize the SIR as ${\rm SIR}=G\cdot W$, where $G=\frac{g(|\varphi_1|) }{g(|\varphi_2|) }$ and $W=\frac{h_1 d_1^{-\alpha_L}}{ h_2 d_2^{-\alpha_L}}$.

\textbf{Lemma 11.} \emph{ Conditioned on  $T_{\phi_A,2}$}, the PDF of $G$ is given by (39) (shown at the bottom of the next page)
\emph{with $g \in [1, g_{max}/g_s]$ and  $\mathbb{P}[T_{\phi_A,2}]  = \frac{1 - e^{-\lambda_{bs} R_L^2 \phi_A} - \lambda_{bs} R_L^2 \phi_A  e^{-\lambda_{bs} R_L^2 \phi_A}}{1-\frac{\Gamma(2,\lambda_{bs} \pi R_L^2)}{\Gamma(2)}}$} .

\textit{Proof.} See Appendix D. \QEDA

\textbf{Lemma 12.} \emph{ Conditioned on $\mathcal{A}_{R_L,2}$,} the PDF of $W$ is given by (40) (shown at the bottom of the next page)
\emph{with $w \in (0,\infty)$}. 
\begin{figure*}[!b]
\hrulefill
\begin{equation}  
\begin{split}
&f_{G|T_{\phi_A,2}}(g)=\frac{1}{\mathbb{P}[T_{\phi_A,2}]}  \int_{g_s}^{\frac{g_{max}}{g}} 
 \frac{5\,( \lambda_{bs} R_L^2 \phi_{3dB})^2}{24\, g\, g_2} \frac{\sqrt{{\rm{log}}_{10}(\frac{g_{max}}{g_2}) {\rm{log}}_{10}(\frac{g_{max}}{g\, g_2})}}{{\rm{ln}}(\frac{g_{max}}{g_2}) {\rm{ln}}(\frac{g_{max}}{g\, g_2})}\, {\rm{exp}}\Big(\frac{\lambda_{bs} R_L^2 \phi_{3dB} \sqrt{10 {\rm{log}}_{10}(g_{max}/g_2)}}{2 \sqrt{3}}\Big) {\rm d} g_2,  
\end{split}
\end{equation}
\begin{equation}  
\begin{split}
&f_{W|\mathcal{A}_{R_L,2}}(w)= \int_0^{\infty}\frac{4 w_2^{-\frac{4}{\alpha_L}-1} (m_s m_x)^{-\frac{2}{\alpha_L}}}{w^{\frac{2}{\alpha_L}+1} \alpha_L^2 R_L^4}  \frac{\Big[\Gamma\Big(\frac{2}{\alpha_L}+m_s\Big)- \Gamma\Big(\frac{2}{\alpha_L}+m_s, \frac{m_s w w_2}{R_L^{-\alpha_L}}\Big)\Big]\Big[\Gamma\Big(\frac{2}{\alpha_L}+m_x\Big)- \Gamma\Big(\frac{2}{\alpha_L}+m_x, \frac{m_x w_2}{R_L^{-\alpha_L}}\Big)\Big]}{\Gamma(m_s)\Gamma(m_x)} {\rm d} w_2,
\end{split}
\end{equation}
\end{figure*}

\textit{Proof.} The PDF of $W_i = h_i r_i^{-\alpha_L}$, $i \in \{1,2\}$ is given by 
\begin{equation} 
\begin{split}
&f_{W_i}(w_i) = \int_{R_L^{-\alpha_L}}^{\infty} \frac{1}{y} f_{d_i^{-\alpha_L}}(y) f_{h_i}(w_i/y) {\rm d} y \\
&= \frac{2 m_u^{-\frac{2}{\alpha_L}}w_i^{-\frac{2}{\alpha_L}-1}\Big[\Gamma\Big(\frac{2}{\alpha_L}+m_u\Big)- \Gamma\Big(\frac{2}{\alpha_L}+m_u, \frac{m_u  w_i}{R_L^{-\alpha_L}}\Big)\Big]}{\alpha_L R_L^2 \Gamma(m_u)}.
\end{split}
\end{equation} 
By setting $W_1= W W_2$, the PDF of $W$ is given by 
\begin{equation} 
f_{W||\mathcal{A}_{R_L,2}}(w) = \int_0^\infty w_2 f_{W_1}(w\, w_2) f_{W_2}(w_2) {\rm d} w_2.
\end{equation}
After applying some simplifications, (40) yields. \QEDA

\textbf{Proposition 2.} \emph{ Conditioned on $\mathcal{A}_{R_L,2}$ and $T_{\phi_A,2}$, the coverage probability in the presence of a dominant angular distance-based interferer under Policy 2 is given by}
 \begin{equation} 
\mathcal{P}_{c}(\gamma)  = 1 - \int_0^\gamma \int_1^{\frac{g_{max}}{g_s}} \frac{1}{g} f_{W|\mathcal{A}_{R_L,2}}(x/g) f_{G||T_{\phi_A,2}}(g) {\rm d} g {\rm d} x.
\end{equation} 

\textit{Proof.}  First, recall that $T_{\phi_A,2} \subseteq \mathcal{A}_{R_L,2}$, which means that $f_{G | \mathcal{A}_{R_L,2}, T_{\phi_A,2}} (g) = f_{G|T_{\phi_A,2}}(g)$.Second, it is easy to argue in this setting that $W$ is conditionally independent of $T_{\phi_A,2}$ when conditioned on  $\mathcal{A}_{R_L,2}$. Therefore,  the PDF of the SIR results directly from expressing the SIR given as a product of two random variables 
\begin{equation} 
f_{SIR} (x)  =  \int_1^{\frac{g_{max}}{g_s}} \frac{1}{g} f_{W|\mathcal{A}_{R_L,2}}(x/g) f_{G|T_{\phi_A,2}}(g) {\rm d} g.
\end{equation} 
Then, $F_{SIR}(\gamma) =  \int_0^\gamma f_{SIR} (x) {\rm d} x$. \QEDA

\subsection{Coverage Probability Under Policy 3}
In this scenario, according to Policy 3 the dominant interferer is the second closest BS in Euclidean distance to the receiver. In this case, the received SIR is given by 
\begin{equation} 
{\rm SIR} = \frac{g_{max} h_1 r_1^{-\alpha_L}}{g(|\varphi_2|) h_2 r_2^{-\alpha_L}} ,
\end{equation}
where $r_1, r_2$ denote the distances of the serving and the dominant interfering BS from the receiver and $|\varphi_2|$ here denotes the absolute angular distance of the interfering node as seen from the receiver. For the dominant interferer scenario, the property of PPP is exploited: If $|\varphi_2| \sim U[0,\pi]$, $|\varphi_2|$ remains uniform in the subset $[0, \phi_A$] i.e., $|\varphi_2| \sim U[0,\phi_A]$. In this scenario, $r_1, r_2$ are dependent. The SIR can be written as ${\rm SIR} = G \cdot W$, where $G = \frac{g_{max} h_1 }{g(|\varphi_2|) h_2 }$ and $W=\frac{ r_1^{-\alpha_L}}{  r_2^{-\alpha_L}}$. 

\textbf{Lemma 13.} \emph{ Conditioned on  $\mathcal{A}_{R_L,2}$,} $f_{G|\mathcal{A}_{R_L,2}}(g)$ \textit{is} 
\begin{equation} 
\begin{split}
&f_{G|\mathcal{A}_{R_L,2}}(g)  =  \frac{5 \sqrt{3}\, \Big(\frac{m_s}{g_{max}}\Big)^{m_s} {m_x}^{m_x} \phi_{3dB}}{6 \sqrt{10}\, {\rm{ln}}(10) \Gamma(m_s) \Gamma(m_x) \phi_A} \\
& \int_0^\infty \int_{\frac{z}{g_{max}}}^{\frac{z}{g_{s}}} \frac{(g\, z)^{m_s-1} e^{-\frac{m_s g z}{g_{max}}-m_x x} x^{m_x -1}}{\sqrt{{\rm{log}}_{10}(x g_{max} /z)}} {\rm d} x\, {\rm d} z. 
\end{split}
\end{equation} 

\textit{Proof.} Along similar lines as in Appendix B,  $f_{g(|\varphi_2|)}(g_2)$ is given by
\begin{equation} 
f_{g(|\varphi_2|)}(g_2)  =  \frac{5 \sqrt{3} \, \phi_{3dB} }{6 \sqrt{10}\, {\rm{ln}}(10)  \phi_A \, g \sqrt{{\rm{log}}_{10}( g_{max} /g)}},  
\end{equation} 
$g \in [g_s, g_{max}]$. Then, $f_{G|\mathcal{A}_{R_L,2}}(g)$ can be obtained by following similar steps with the proof for Lemma 12. \QEDA

\emph{ \textbf{Lemma 14.}  Conditioned on  $\mathcal{A}_{R_L,2}$, $f_{W|\mathcal{A}_{R_L,2}}(w)$ is }
\begin{equation} 
\begin{split}
& f_{W|\mathcal{A}_{R_L,2}}(w)= \frac{\Big(\frac{1}{w}\Big)^{\frac{2 + \alpha_L}{\alpha_L}}}{\alpha_L}\\
&  \times \Big(\frac{3 {\rm{erf}}(\sqrt{\pi \lambda_{bs}})}{2 \lambda_{bs}} - 2 (1+R_L^2 \lambda_{bs} \pi) e^{- R_L^2 \lambda_{bs} \pi} - e^{-\lambda_{bs} \pi}   \Big) 
\end{split}
\end{equation}
\emph{$w \in [1, \infty)$, and ${\rm{erf}}(\cdot)$ denotes the error function \cite[eq. (8.250.1)]{Ryzhik}}.

\textit{Proof.} The joint PDF of $f_{r_1,r_2 | \mathcal{A}_{R_L,2}}(r_1,r_2)$ is first derived. By exploiting theory of order statistics  and following similar procedure as in Appendix A,  the joint PDF of $f_{r_1,r_2 | \mathcal{A}_{R_L,2}}(r_1,r_2)$ is given by
\begin{equation} 
f_{r_1,r_2 | \mathcal{A}_{R_L,2}}(r_1,r_2) = \frac{4 (\pi \lambda_{bs})^2 r_1 r_2 e^{- \lambda_{bs} \pi r_2^2}}{1-\frac{\Gamma(2,\lambda_{bs} \pi R_L^2)}{\Gamma(2)}}, 
\end{equation}
for $r_1 \in [0,R_L], r_2 \in [r_1,R_L]$. Then, the PDF of $W$ is obtained by first obtaining the PDF of $v=r_1/r_2$ and then through the transformation $W=v^{-\alpha_L}$. \QEDA 

\textbf{Proposition 3.} \emph{ Conditioned on $\mathcal{A}_{R_L,2}$, the coverage probability in the presence of a dominant interferer under Policy 3 is given by}
 \begin{equation} 
\mathcal{P}_{c}(\gamma)  = 1 - \int_0^\gamma \int_1^{\infty} \frac{1}{g} f_{W|\mathcal{A}_{R_L,2}}(x/g) f_{G|\mathcal{A}_{R_L,2}}(g) {\rm d} g {\rm d} x.
\end{equation} 

\textit{Proof.} The proof follows the same steps as the proof in Proposition 1. \QEDA

\section{Results and Discussion}
In this section, numerical results are presented to evaluate and compare the performance achieved in a dense mmWave cellular network under different association policies. The accuracy of the analytical results is verified by comparing them with the empirical results obtained from Monte-Carlo simulations. For all numerical results, the following parameters have been used unless stated otherwise: $R_L = 75$ meters, $\alpha_L= 2$, $\alpha_N= 3.5$, $f_c=26.5$ GHz as in \cite{Baccelli} and $R=100$ meters. As per the 3GPP specifications $p = 45$ dBm, $\sigma^2 = -74$ dBm,  $\lambda_{bs}= 0.0008$ BSs/$m^2$ and $m_u= 2$. The receiver is assumed to be equipped with a directional antenna with 4 sectors\footnote{Very large codebooks can be considered at the receiver, thereby resulting in a huge number of narrow beams. However, \cite{3GPPmodel} states that the SSB-based requirements upper-bound the number of RX beams to 8. Besides, to avoid the high overhead and complexity issues associated with wide spatial domain coverage with a huge number of very narrow beams, on which CSI-RSs are transmitted, it is reasonable to consider only subsets of those beams, usually based on the locations of the active receiver. This is also important for UE power consumption considerations.}, i.e., $m= 2$ and $\phi_{3dB} = \pi/2$.

\begin{figure}[!t]
    \centering
    \includegraphics[keepaspectratio,width= \linewidth]{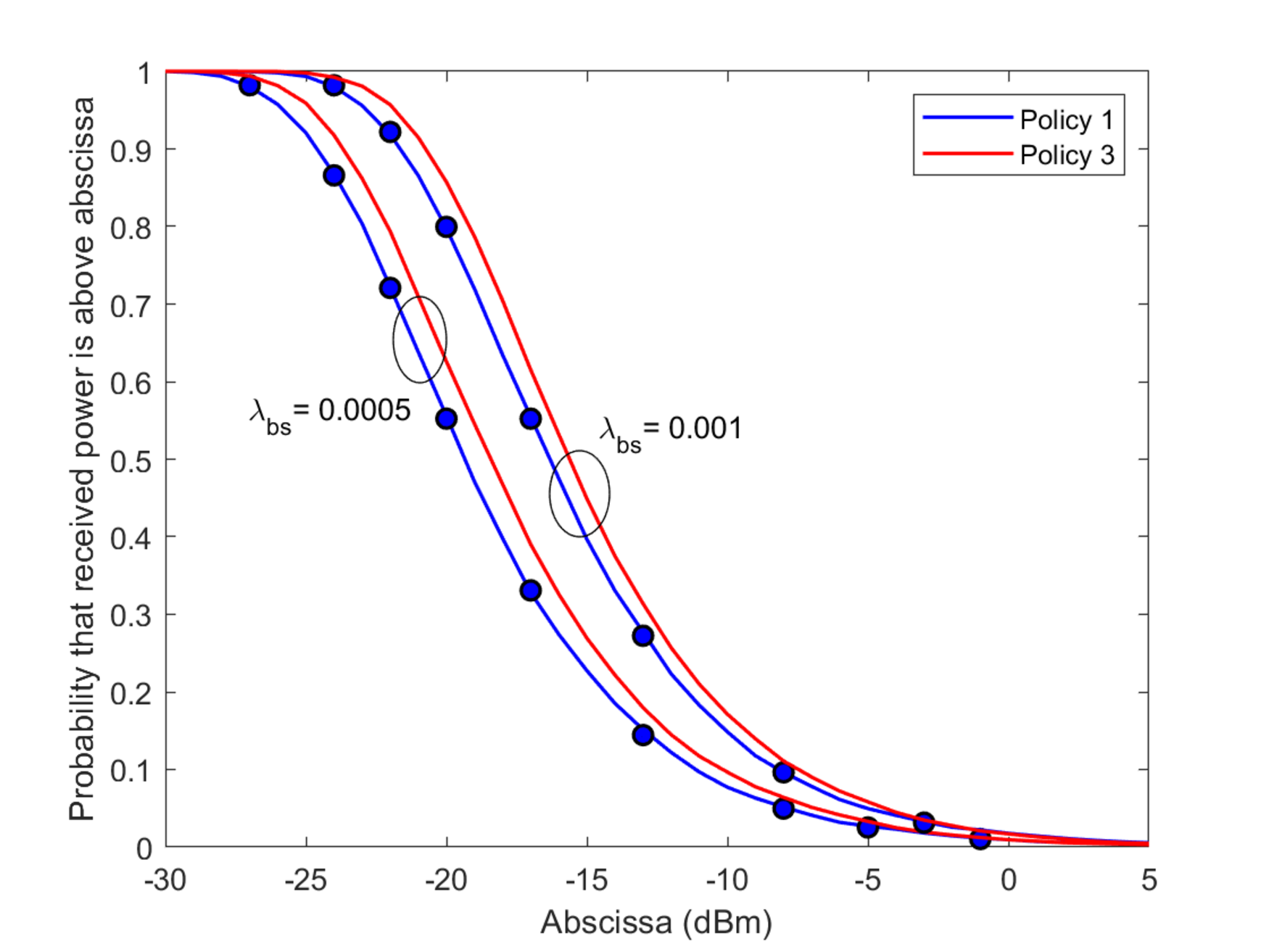}
       \caption{CCDF of received power under Policy 1 and Policy 3, for different values of $\lambda_{bs}$. Markers denote the analytical results.}
\end{figure}

\begin{figure}[!t]
    \centering
    \includegraphics[keepaspectratio,width= \linewidth]{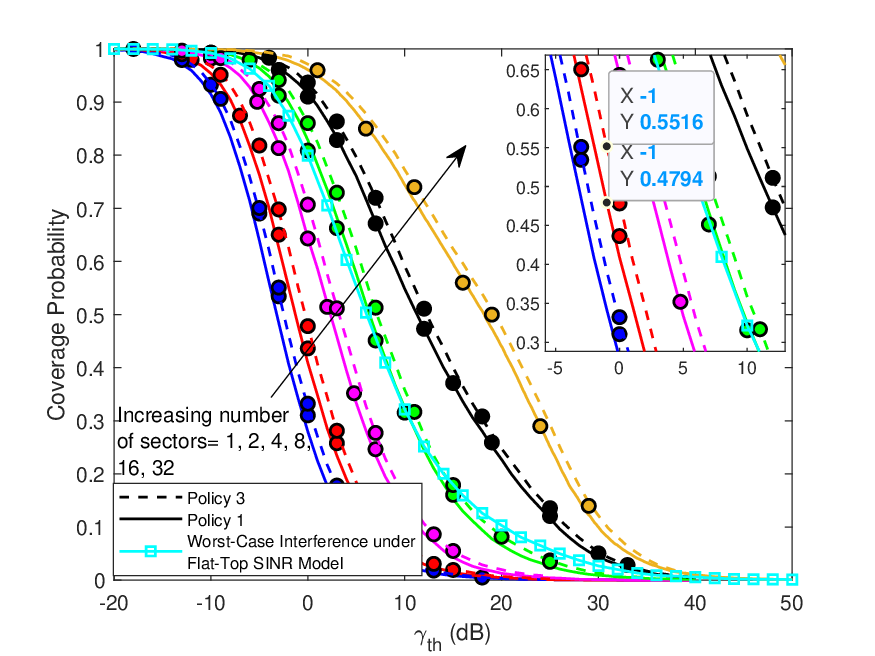}
       \caption{ Coverage probability versus $\gamma_{th}$ under Policy 1 and Policy 3, for different number of sectors. Markers denote the analytical results. Simulation results under a flat-top antenna pattern are also presented for completeness.} 
\end{figure}

Fig. 4 presents the CCDF of the maximum received power under Policy 1, defined as in (20) and evaluated through Lemma 3, for different values of $\lambda_{bs}$. To address and highlight the impact of realistic association schemes in the performance of mmWave networks, the results are compared through simulations to the received power of Policy 3, defined as $S_r = g_{max} r_1^{-\alpha_L}$, where the receiver is associated to the nearest LOS BS and no misalignment error exists. It is observed that Policy 3 overestimates the received power compared to the maximum received power achieved under Policy 1, especially for small values of the abscissa. Interestingly, notice that the misalignment error exists even in denser mmWave networks. Consequently, associating to the nearest LoS BS is not realistic, even in dense mmWave networks.

Fig. 5 compares the coverage probability versus $\gamma_{th}$ under Policy 1 and Policy 3 for several numbers of beams/sectors produced by the receiver antenna. It is observed that the coverage performance of the network under Policy 3 is overestimated and the difference in coverage performance is not negligible  (e.g. for $\gamma_{th} = -1$ dB, the coverage probability under Policy 3 is 0.55, while the coverage probability under Policy 1 is 0.48). Note that in Policy 1, the receiver performs exhaustive scanning in each sector to select the BS that provides the maximum power. The number of beams/sectors of the receiver's antenna affects both the desired received power and the interference power falling within the 3dB beamwidth of the receiver's antenna pattern, while in Policy 3 the number of sectors only determines the interference power. Indeed, by increasing the number of sectors in Policy 1, the receiver minimizes the misalignment error. At the same time, the receiver's beams become highly-directional, thus decreasing the interference power, and the coverage performance approaches the corresponding one under Policy 3.  The cyan curve depicts coverage performance for the worst-case interference scenario under the flat-top antenna pattern approximation. It is observed that the coverage performance under the flat-top pattern approximation is  overestimated w.r.t. the performance under Policy 1, especially in the higher SINR regime. Therefore, the joint effect of the approximation of the flat-top antenna pattern and perfect alignment on the coverage performance under Policy 1, is captured.

\begin{figure}[!t]
    \centering
    \includegraphics[keepaspectratio,width= \linewidth]{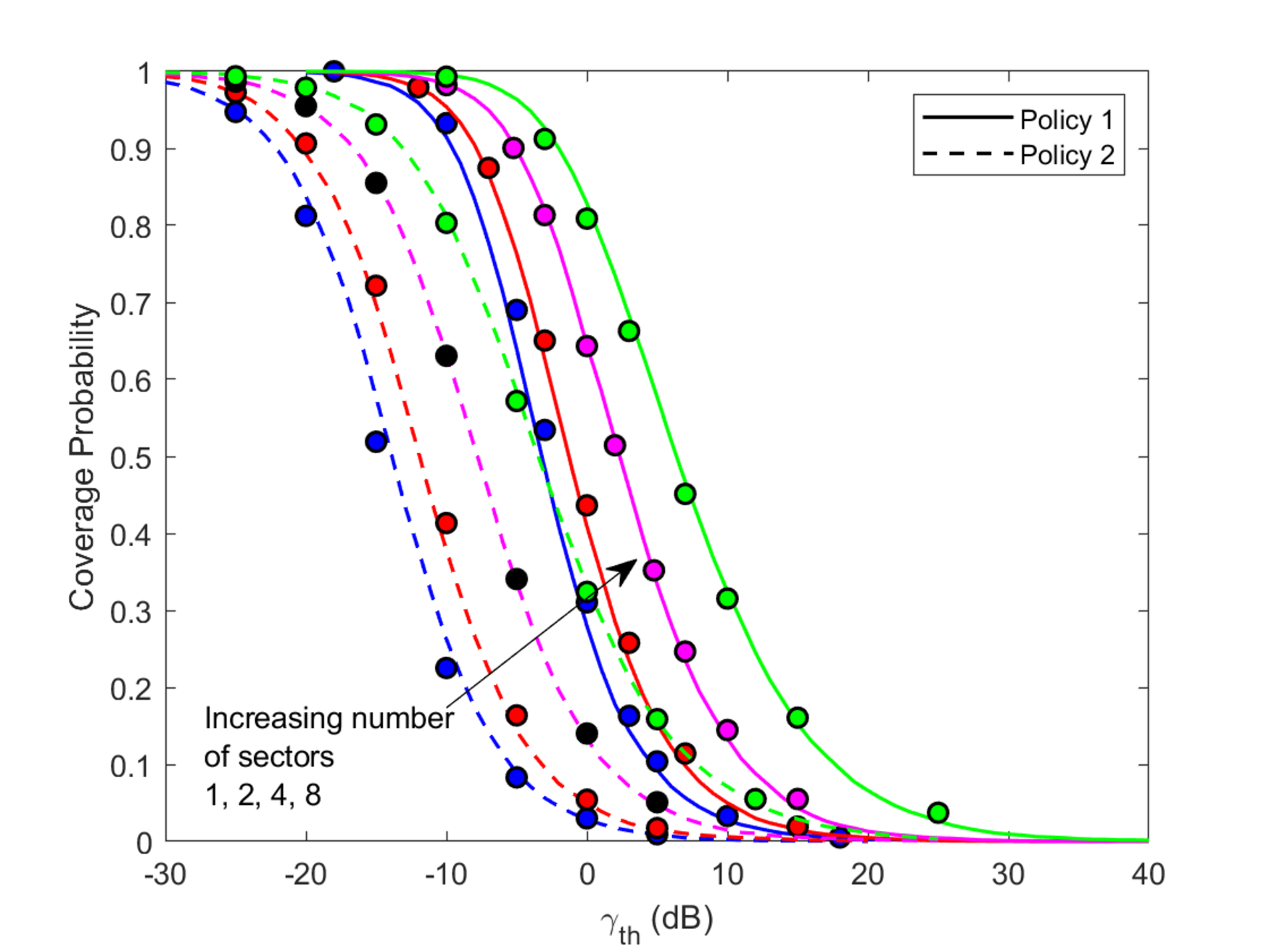}
       \caption{ Coverage probability versus $\gamma_{th}$ under Policy 1 and Policy 2, for several numbers of sectors. Markers denote the analytical results.}
\end{figure}

\begin{figure}[!t]
    \centering
    \includegraphics[keepaspectratio,width= \linewidth]{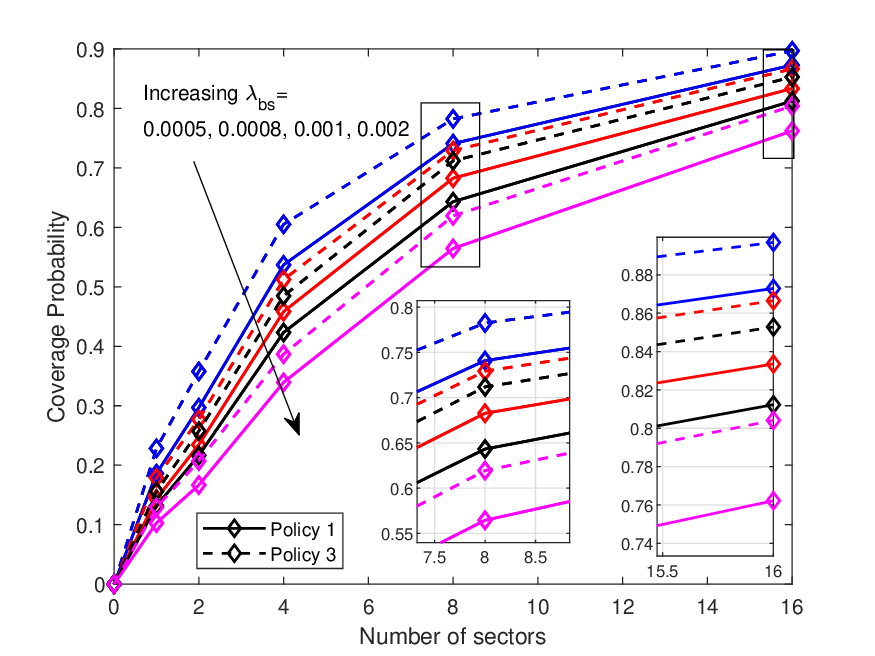}
       \caption{ Coverage probability versus number of sectors under Policy 1 and Policy 3, for different values of $\lambda_{bs}$ and $\gamma_{th}= 3$dB.}
\end{figure}

Fig. 6 compares the coverage probability versus $\gamma_{th}$ under Policy 1 and Policy 2 when the receiver's antenna is equipped with different number of sectors. In both of the association schemes, the receiver performs maximum power-based and angular distance-based exhaustive scanning, respectively to select the serving BS. It is observed that when only angular distance-based scanning is performed, the coverage performance is significantly underestimated. This result indicates that: i) Considering merely angular distance-based criteria for estimating the coverage performance is inaccurate ii) The path-loss plays a crucial role in the performance of mmWave networks.

Fig. 7 compares the coverage probability versus the number of sectors of the receiver's antenna under Policy 1 and Policy 3, derived analytically in Theorem 1 and Theorem 3, respectively, for different values of $\lambda_{bs}$ and target $\gamma_{th} =3$ dB. One can observe that  the coverage performance under ideal baseline scenario overestimates the corresponding one under Policy 1, which accounts for the misalignment error, for all BSs' deployment densities. With the increase of the number of sectors, the UE's beams in Policy 1 become more directional which alleviates beam misalignment effects and the coverage performance approaches the corresponding one under the ideal baseline scenario. This also highlights the need for explicitly modeling of misalignment error for realistic antenna patterns, even a small misalignment error will impact the performance and hence, the angular distances become crucial in realistic performance analysis of mmWave networks.

\begin{figure}[!t]
    \centering
    \includegraphics[keepaspectratio,width= \linewidth]{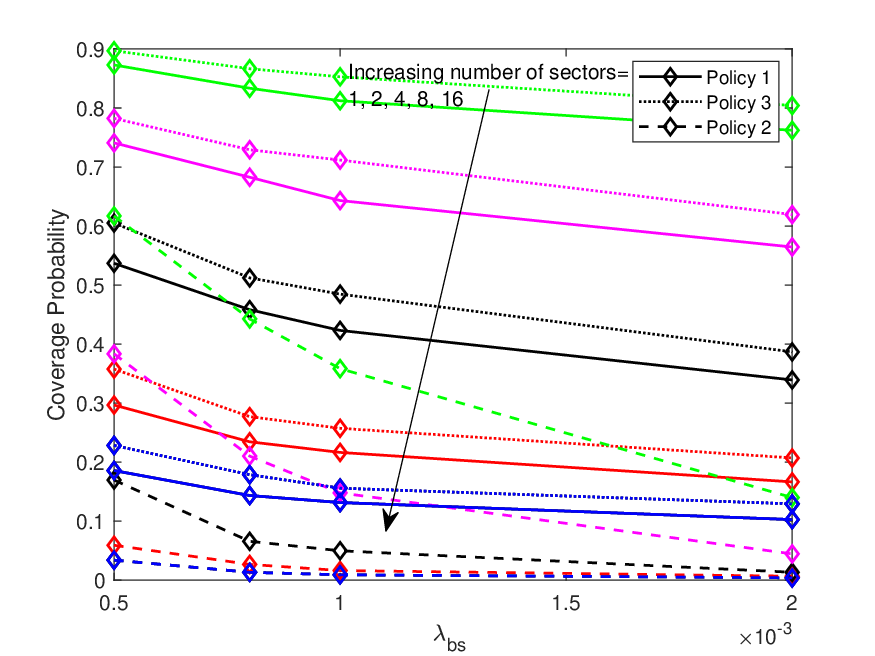}
       \caption{ Coverage probability versus $\lambda_{bs}$ for different number of sectors under the three policies, for  $\gamma_{th}= 3$dB.}
\end{figure}

 Fig. 8 shows the coverage probability versus $\lambda_{bs}$ for different number of sectors of the receiver’s antenna, under the three policies. It is observed that Policy 3 slightly overestimates the coverage performance of the network for all number of sectors. On the other hand, it is clearly observed that Policy 2 significantly underestimates the network's coverage performance. This result clearly confirms that path-loss cannot be ignored during the determination of the serving BS.

Fig. 9 shows the coverage probability versus $\gamma_{th}$ under  Policy 1, Policy 2 and Policy 3 for the dominant interferer. Moreover, the coverage performance under Policy 1 and aggregate interference is depicted, when the receiver's antenna is equipped with different number of sectors. Quite interestingly, it is observed that the coverage performance under Policy 2 with a single dominant interferer approximates the performance under Policy 1 with aggregate interference, especially when the number of sectors is small. In this case, the angular distance-based criterion results in realistic network performance. On the other hand, the performance under Policy 3 in the presence of a single dominant interferer clearly overestimates  the corresponding one under Policy 1, which leads to the following system-level outcome: \emph{In a LOS ball of a mmWave network under beam misalignment error at the receiver, by attaching to the closest  BS in angular distance and considering the dominant interferer as the closest BS in angular distance w.r.t. the line of communication link, results in a more accurate approximation of the coverage performance compared to the policy of attaching to the closest LOS BS and considering the dominant interferer as the second nearest BS.}  Notably, the performance of Policy 2 under dominant interferer approach yields a better approximation of the network's coverage than the corresponding performance under Policy 1 under dominant interferer approach.

\begin{figure}[!t]
    \centering
    \includegraphics[keepaspectratio,width= \linewidth]{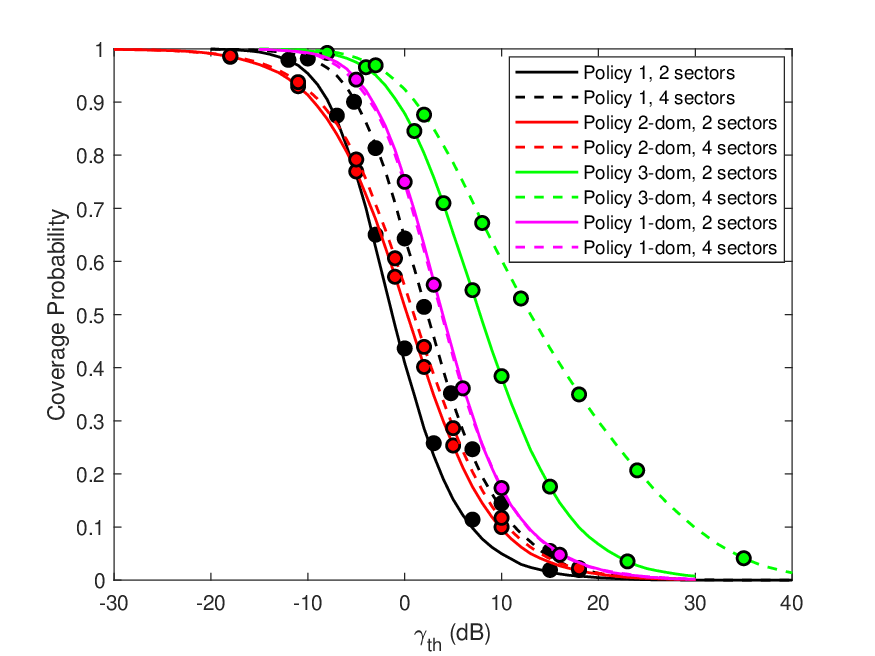}
       \caption{ Coverage probability versus number of sectors under Policy 1 and Policy 3, for different values of $\lambda_{bs}$ and $\gamma_{th}= 3$dB.}
\end{figure}

\begin{figure*}[!htbp]
  \begin{subfigure}[t]{.33\linewidth}
  \includegraphics[trim=20 15 20 15,clip,width=\linewidth]{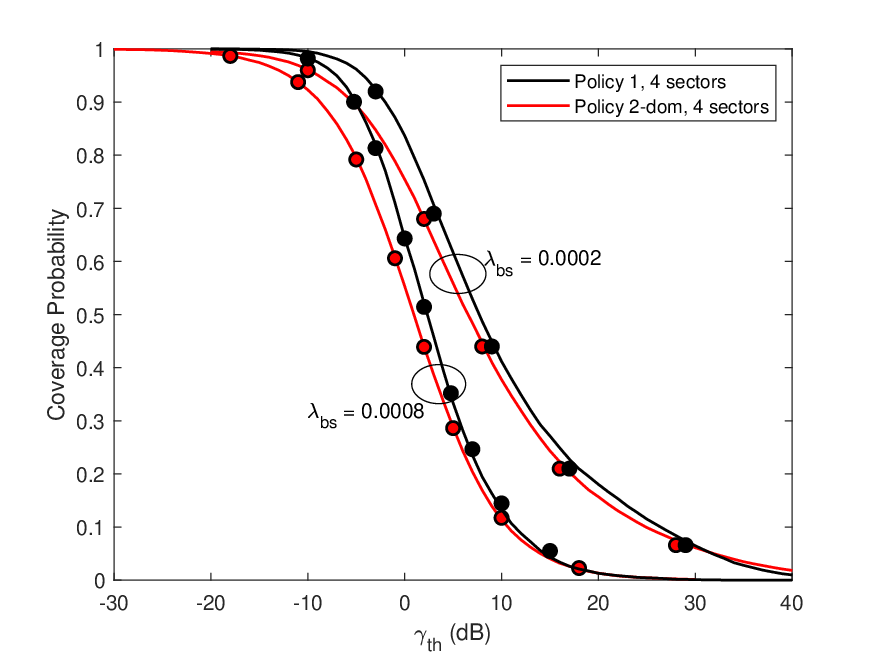}
    \caption{Different values of $\lambda_{bs}$.}%
\label{a}
  \end{subfigure}\hfil
  \begin{subfigure}[t]{.33\linewidth}
    \includegraphics[trim=20 15 20 15,clip,width=\linewidth]{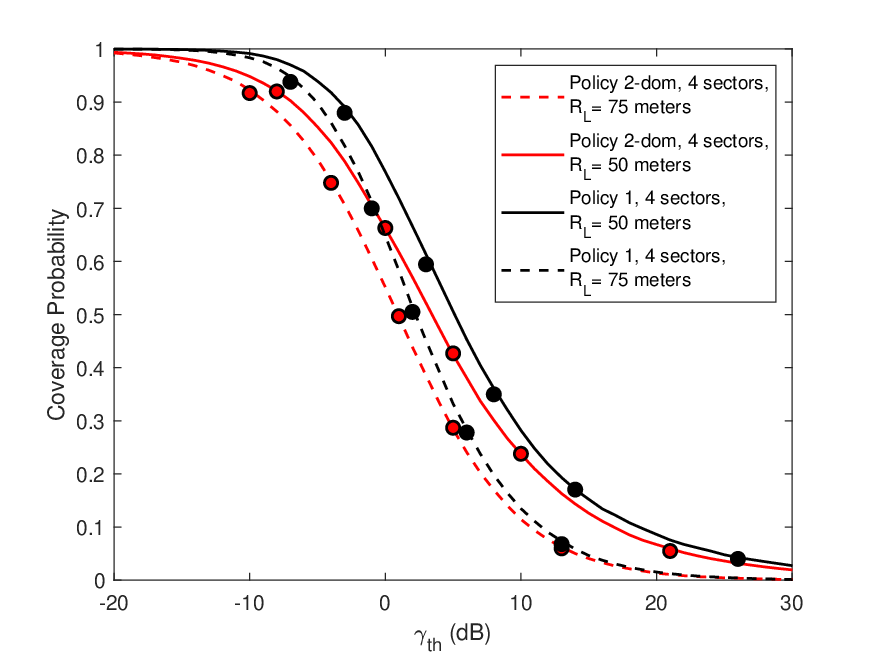}%
    \caption{ Different values of $R_L$}%
  \end{subfigure}\hfil
  \begin{subfigure}[t]{.33\linewidth}
    \includegraphics[trim=20 15 20 15,clip,width=\linewidth]{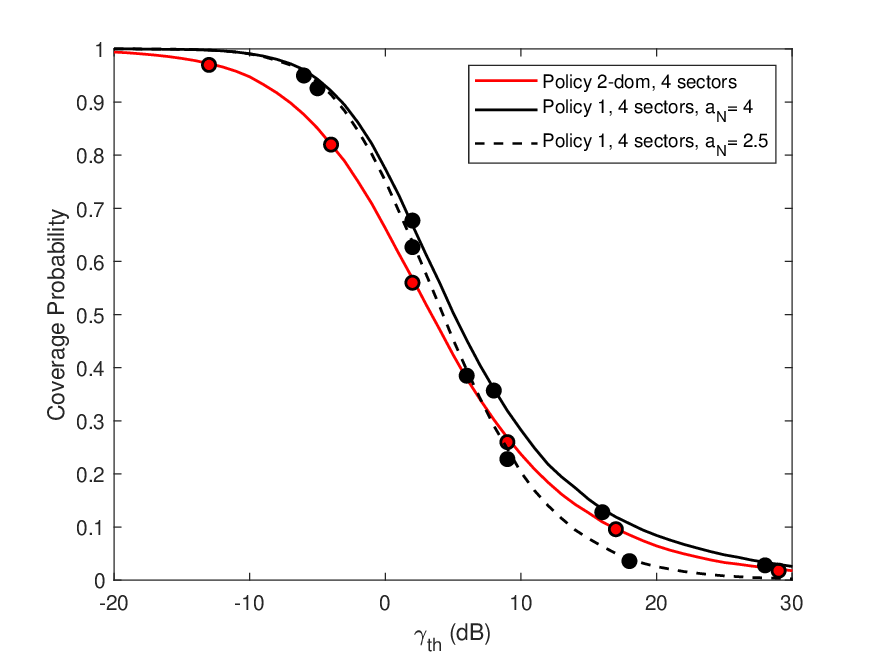}%
    \caption{Different values of $\alpha_N$. }%
  \end{subfigure}%
   \caption{Coverage probability versus $\gamma_{th}$ under Policy 1 and Policy 2 dominant
interferer approach. Markers denote the analytical results.  }
\end{figure*}

 Fig. 10 shows the coverage probability versus $\gamma_{th}$ under Policy 1 and Policy 2 dominant interferer approaches. This figure investigates different network setups for which the performance under Policy 2 dominant interferer approach yields a close approximation for the network's coverage performance. In Fig. 10a, coverage probability versus $\gamma_{th}$ under Policy 1 and Policy 2 dominant interferer approach for different values of $\lambda_{bs}$, is illustrated. It is observed that the performance under Policy 2 dominant interferer approach closely approximates the network's coverage performance for both dense and sparse mmWave networks, especially in the higher SINR regime. In Fig. 10b, coverage probability versus $\gamma_{th}$ under Policy 1 and Policy 2 dominant interferer approach for different values of $R_L$, is illustrated. In a similar vein, it is observed that the performance under Policy 2 dominant interferer approach is a good approximation of the network's coverage performance for both values of $R_L$, especially in the higher SINR regime. Finally, in Fig. 10c, coverage probability versus $\gamma_{th}$ under Policy 1 and Policy 2 dominant interferer approach for different values of $a_N$ in Policy 1, is illustrated. Clearly, the performance under Policy 2 dominant interferer approach yields a close approximation of the network's coverage performance for both values of $a_N$, especially as the interference power outside the LOS ball becomes weaker. 

 \emph{Remark 2.} \textit{The received power under Policy 2 tends to zero for infinitely large LOS ball radius. However, infinite radius is practically impossible to occur in realistic mmWave networks where typical values are smaller than 100 meters \cite{Baccelli}.}

\section{Conclusions}
In this work, a novel stochastic geometry framework was proposed for mmWave cellular networks to address the role of the angular distances in a maximum power based association policy under a realistic beam management procedure. This necessitated the inclusion of both angular and Euclidean distances in the analysis of cell association, which is a key novelty of this paper. Three association policies were considered for comparison. Subsequently, coverage probability analysis was conducted for the three policies and exact-form expressions were derived. Subsequently, the definition for the dominant interferer in mmWave networks is reconsidered and performance analysis is conducted under the dominant interferer approach  for \textit{all} policies. As a key system-level insight, it was shown that considering merely Euclidean distance-based policy for determining both the serving and the dominant interfering BS, even in a LOS ball in mmWave networks, is inaccurate. Moreover, the impact of angular distances in the misalignment error was addressed. Indeed, by considering realistic  patterns for capturing misalignment, even small errors will impact the received signal power and hence angular distances play a key role in realistic performance analysis of mmWave networks.

\appendices
\section{Proof of  Lemma 2}
 Consider the event $\mathcal{A}_{r,2}$. Following the properties of a PPP, the spatial locations of the $K$ points in $b(o,r)$ form a finite PPP and the number of points falling in $b(o,r)$ follows a Poisson distribution with mean $\lambda |b(o,r)|$. The set of the unordered absolute angular distances $\{|\phi_k|\}_{k=1:K}$ for the $k$th angular distance of $k$th point, is uniformly distributed on $[0,\pi]$, i.e., $|\phi_k| \sim U[0,\pi]$. By order statistics \cite{orderstatistics},  let $Z_{r,k}$ define the event that $k$ BSs exist in $b(o,r)$.  The joint PDF of the smallest two random variables $X_1, X_2$, given $Z_{r,K}$, is given by $ f_{X_1, X_2 }(x,y\,|\,Z_{r,K}) =  K (K-1) [1-F_{X_1}(y)]^{K-2} f_{X_1}(x) f_{X_1}(y)$.
Let $X_1=|\varphi_1|$,  $X_2=|\varphi_2|$. Then, $f_{|\varphi_1|,|\varphi_2|}(|\varphi_1|,|\varphi_2| \,|\, Z_{r,K})$ is given by $ f_{|\varphi_1|,|\varphi_2| }(|\varphi_1|,|\varphi_2| \,|\, Z_{r,K}) = \frac{K(K-1)}{\pi^2}  \Bigg(1-\frac{|\varphi_2|}{\pi}\Bigg)^{K-2}$,
and corresponding CDF is given by 
\begin{equation} \label{A3}  
\begin{split}
&F_{|\varphi_1|,|\varphi_2|}(|\varphi_1^0|,|\varphi_2^0| \,|\, Z_{r,K})\\
&=  \int_{0}^{|\varphi_1^0|} \int_{|\varphi_1|}^{|\varphi_2^0|} f_{|\varphi_1|,|\varphi_2|}(|\varphi_1|,|\varphi_2| | Z_{r,K} ) \, {\rm d} |\varphi_2| \, {\rm d} |\varphi_1| \\
& = \frac{K}{\pi^K} \Big( \frac{\pi^K - (\pi-|\varphi_1^0|)^K - K |\varphi_1^0|  (\pi-|\varphi_2^0|)^{K-1}}{K}\Big).
\end{split}
\end{equation}  
Then, for the PPP case over a finite region, $F_{|\varphi_1|,|\varphi_2| | \mathcal{A}_{r,2}}(|\varphi_1^0|,|\varphi_2^0|)$ can be obtained as
\begin{equation}  \label{A4}
\begin{split}
 &F_{|\varphi_1|,|\varphi_2| | \mathcal{A}_{r,2}}(|\varphi_1^0|,|\varphi_2^0|) \\
 &= \frac{\mathbb{P}[|\varphi_1|<|\varphi_1^0|,|\varphi_2|<|\varphi_2^0|,  \mathcal{A}_{r,2}]}{\mathbb{P}[ \mathcal{A}_{r,2}]}.  
 \end{split}
\end{equation} 
Now, the probability that the event \{$|\varphi_1|<|\varphi_1^0|,|\varphi_2|<|\varphi_2^0|,  \mathcal{A}_{r,2}$\} occurs, can be obtained by averaging over $ \mathcal{A}_{r,2}$, that is
\begin{equation} \label{A5} 
\begin{split}
&\mathbb{P}[|\varphi_1|<|\varphi_1^0|,|\varphi_2|<|\varphi_2^0|,  \mathcal{A}_{r,2}]  \\
&= \sum_{k=2}^{\infty} F_{|\varphi_1|,|\varphi_2| }(|\varphi_1^0|,|\varphi_2^0| \,|\, Z_{r,k}) \,\frac{(\lambda \pi r^2)^k}{k!} e^{-\lambda \pi r^2} \\
&= 1 - e^{-\lambda r^2 |\varphi_1^0|} - \lambda r^2 |\varphi_1^0|  e^{-\lambda r^2 |\varphi_2^0|}.
\end{split}
\end{equation}
Next, $\mathbb{P}[ \mathcal{A}_{r,2} ] = 1-\frac{\Gamma(2,\lambda \pi r^2)}{\Gamma(2)}$. By substituting in (52), $F_{|\varphi_1|,|\varphi_2| | \mathcal{A}_{r,2}}(|\varphi_1^0|,|\varphi_2^0|)$ is given by
\begin{equation}  \label{A6}
 F_{|\varphi_1|,|\varphi_2| | \mathcal{A}_{r,2}}(|\varphi_1^0|,|\varphi_2^0|) = \frac{1 - e^{-\lambda r^2 |\varphi_1^0|} - \lambda r^2 |\varphi_1^0|  e^{-\lambda r^2 |\varphi_2^0|}}{1-\frac{\Gamma(2,\lambda \pi r^2)}{\Gamma(2)}}.  
\end{equation}
Finally,  $f_{|\varphi_1|,|\varphi_2|| \mathcal{A}_{r,2} }(|\varphi_1|,|\varphi_2|) $ can be obtained after differentiating \eqref{A6}  w.r.t $|\varphi_1^0|,|\varphi_2^0|$ as $ f_{|\varphi_1|,|\varphi_2| | \mathcal{A}_{r,2}}(|\varphi_1|,|\varphi_2|) = \frac{\partial^2 F_{|\varphi_1|,|\varphi_2| | \mathcal{A}_{r,2}}(|\varphi_1^0|,|\varphi_2^0|) }{\partial |\varphi_1^0| \partial |\varphi_2^0|}$,
which directly results to (\ref{eq5}).

\section{Proof of Lemma 3}
The PDF of $S_{x}$ is first derived. Since $\varphi_r \sim U[0,\frac{\phi_{3dB}}{2}]$, $g(\varphi^x) = 10^{\frac{G_{max}-12\big( \frac{\varphi^x}{\phi_{3dB}}\big)^2}{10}}$. 
Then, $g(\varphi^x)$ is a function of the random variable $\varphi^x$. Building on $\varphi^x$ and applying successive change of variables, the PDF $f_{g(\varphi^x)}(g)$ is given by $f_{g(\varphi^x)}(g) = \frac{1}{ln(10) x}\frac{10}{12 \sqrt{\frac{G_{max}-10 log(x)}{12}}}, \quad x \in [g_{3dB},g_{max}]$.
Subsequently, the PDF of $r_x^{-\alpha_L}$ is expressed in terms of the corresponding CDF as 
\begin{equation}\label{B2}
\begin{split}
&\mathbb{P}[r_x^{-\alpha_L} \leq x ] = \mathbb{P}\Big[r_x^{\alpha_L} \geq \frac{1}{x} \Big] = 1-\mathbb{P}\Big[ r_x \leq \Big(\frac{1}{x}\Big)^{\frac{1}{\alpha_L}}\Big]\\
& = 1-F_{r_x}\Big(\Big(\frac{1}{x}\Big)^{\frac{1}{\alpha_L}}\Big), \quad F_{r_x}(r) = \frac{r^2}{R_L^2}.
\end{split}
\end{equation}
Now, $f_{r_x^{-\alpha_L}}(x)$ is obtained after differentiating \eqref{B2}  w.r.t the appropriate range of $x$, that is, $f_{r_x^{-\alpha_L}}(x) = \frac{2 \big(\frac{1}{x} \big)^{\frac{\alpha_L+2}{\alpha_L}}}{\alpha_L R_L^2},  x \in [R_L^{-\alpha_L}, \infty).$ The PDF and the CDF of $S_{x}$ can now be written as 
\begin{equation}\label{B4}  
f_{S_{x}}(w) =   \int_{g_{3dB}}^{\psi(w)}\frac{1}{ x} f_{g(\varphi^x)}(x) f_{r_x^{-\alpha_L}}\Big(\frac{w}{x}\Big) {\rm d} x,
\end{equation}
\begin{equation}  \label{B5}
F_{S_{x}}(w_0) =  \int_{g_{3dB}R_L^{-\alpha_L}}^{w_0} \int_{g_{3dB}}^{\psi(w)}\frac{1}{ x} f_{g(\varphi^x)}(x) f_{r_x^{-\alpha_L}}\Big(\frac{w}{x}\Big) {\rm d} x {\rm d} w,
\end{equation}
 The PDF of $S = \underset{x \in (\Psi_{bs} \cap \mathbf{b}(\mathbf{o},R_L))}{\operatorname{max}} \{S_{x}\} $ can now be obtained by exploiting results from Order Statistics \cite{orderstatistics}.  However, $S$ is meaningfully defined conditioned on $\mathcal{A}_{R_L,1}$. Accordingly, let $Z_{R_L,k}$, with $k \geq 1$, define the event that $k$ BSs exist in $ \mathbf{b}(\mathbf{o},R_L)$. Then, from the definition of $\Psi_{bs} \cap \mathbf{b}(\mathbf{o},R_L)$, $\mathbb{P}[\mathcal{A}_{R_L,k}] = e^{-\lambda_{bs} \pi R_L^2} (\lambda_{bs} \pi R_L^2)^k/k!$.  Given $Z_{R_L,k}$ and since the elements of $S_{x}$ are i.i.d., the probability that $S \leq s_0$ is given by $\mathbb{P}[S \leq s_0|Z_{R_L,k}]=F_{S}(s_0|Z_{R_L,k})=   (F_{S_{x}}(w))^k$. Now,  in a similar vein to the proof presented in Appendix A, $F_{S|\mathcal{A}_{R_L,1}}(s_0)$ can be obtained as $F_{S|\mathcal{A}_{R_L,1}}(s_0) = \frac{\mathbb{P}[S \leq s_0, \mathcal{A}_{R_L,1}]}{\mathbb{P}[\mathcal{A}_{R_L,1} ]}$.
Accordingly, $\mathbb{P}[S \leq s_0, \mathcal{A}_{R_L,1}]$ can be derived by averaging $\mathbb{P}[S \leq s_0|\mathcal{A}_{R_L,k}]$ over $\mathcal{A}_{R_L,k}$, that is 
\begin{equation}\label{B7}
\begin{split} 
& \mathbb{P}[S \leq s_0, \mathcal{A}_{R_L,1}] \\
& \myeqa \sum_{k=1}^{\infty} (F_{S_{x}}(w))^k e^{-\lambda_{bs} \pi R_L^2} \frac{(\lambda_{bs} \pi R_L^2)^k}{k!} \\
& \myeqb e^{\lambda_{bs} \pi R_L^2 \big(1-F_{S_{x}}(s_0)\big)} - e^{-\lambda_{bs} \pi R_L^2 }, \quad s_0 \in [w_{min},\infty),
\end{split}
\end{equation}
where (a) follows after averaging over $k$ and (b) follows from $\sum_{k=1}^{\infty} \frac{a^k b^k}{k!} \myeqdef e^{a b} -1$. Next, $\mathbb{P}[\mathcal{A}_{R_L,1} ] = 1 - e^{- \lambda \pi R_L^2}$. By substituting all the above in the definition of $F_{S|\mathcal{A}_{R_L,1}}(s_0)$, $F_{S|\mathcal{A}_{R_L,1}}(s_0)$ yields. Finally, the PDF $ f_{S|\mathcal{A}_{R_L,1}}(s_0)$ is obtained directly as $f_{S|\mathcal{A}_{R_L,1}}(s_0) = \frac{d F_{S|\mathcal{A}_{R_L,1}}(s_0)}{d s_0} = \frac{ \lambda_{bs} \pi R_L^2 \,f_{S_{x}}(s_0) \,e^{\lambda_{bs} \pi R_L^2 \big( F_{S_{x}}(s_0) - 1 \big) }}{1 - e^{- \lambda \pi R_L^2}} $. 

\begin{figure*}[!hb]
\hrulefill
\begin{equation}  \label{C2}
\begin{split}
&\mathcal{L} _{I}(s|S_{th},\phi_0^r)=\mathbb{E}_{\Psi_{bs}^{!}}[e^{-sI}]=\mathbb{E}_{\Psi_{bs}^{!}}\Big[e^{-s \sum_{x \in \Psi_{bs}^{!}} p h_x g_{max} g_{3gpp}(|\phi_0^r-\phi_x|) l(\|x\|)}\Big|S_{th},\phi_0^r\Big] \\
&\myeqa \mathbb{E}_{\Psi_{bs}^{!}}\Big[\prod_{x \in \Psi_{bs}^{!}} \Big(1+\frac{s\, p \,g_{max}\, g_{3gpp}(|\phi_0^r-\phi_x|)  l(\|x\|)}{m_x} \Big)^{-m_x} \Big|S_{th},\phi_0^r \Big] \\
&\myeqb {\rm{exp}}\Bigg( -\lambda_{bs} \iint_\mathbf{\Omega} \Big(1- \Big(1+\frac{s\, p\,  g_{max}\, g_{3gpp}(|\phi_0^r-\phi_x|)\,K r_x^{-\alpha_L}}{m_x} \Big)^{-m_x}\Big) r_x {\rm d} \phi_x {\rm d} r_x \Bigg) \\
& \times  {\rm{exp}}\Bigg( -\lambda_{bs} \iint_{\mathbf{\Omega^{'}}} \Big(1- \Big(1+\frac{s\, p\,  g_{max}\, g_{3gpp}(|\phi_0^r-\phi_x|)\,K r_x^{-\alpha_N}}{m_x} \Big)^{-m_x}\Big) r_x {\rm d} \phi_x {\rm d} r_x  \Bigg) \\
&\myeqc {\rm{exp}}\Bigg(-\lambda_{bs} \int_{r_x^{min}}^{R_L} \int_{0}^{2 \pi} \Big(1-\Big(1+\frac{s\, p\, g_{max}\, g_{3gpp}(|\phi_0^r-\phi_x|) K r_x^{-\alpha_L}}{m_x} \Big)^{-m_x}  \Big) \Bigg) r_x {\rm d} \phi_x {\rm d} r_x \\
&\times  {\rm{exp}}\Bigg(-\lambda_{bs} \int_{R_L}^{r_x^{max}} \int_{0}^{2 \pi} \Big(1-\Big(1+\frac{s\, p\, g_{max}\, g_{3gpp}(|\phi_0^r-\phi_x|) K r_x^{-\alpha_N}}{m_x} \Big)^{-m_x}  \Big) \Bigg) r_x {\rm d} \phi_x {\rm d} r_x, 
\end{split}
\end{equation}
\end{figure*}

\section{Proof of Lemma 4}
The Laplace transform of $I$ conditioned on $S$ and $\phi_0^r$ is given by (59) (shown at the bottom of the next page) where (a) follows from the  moment generating function (MGF) of $h_x$, (b) follows from the probability generating functional (PGFL) of the PPP and  from independence between the locations of LOS and NLOS interfering BSs, and (c) follows from integration over the set of the region $\mathbf{\Omega}$ and  $\mathbf{\Omega^{'}}$.

\section{Proof of Lemma 11}
 Since $|\varphi_1|, |\varphi_2|$ are dependent and conditioned on  $T_{\phi_A,2}$, the PDF $f_{|\varphi_1|,|\varphi_2| | | T_{\phi_A,2}}(|\varphi_1|,|\varphi_2|)$ of $|\varphi_1|, |\varphi_2|$ is given as
\begin{equation} \label{D1} 
\begin{split}
 &f_{|\varphi_1|,|\varphi_2| | T_{\phi_A,2}}(|\varphi_1|,|\varphi_2| ) \\
 &= \frac{1}{\mathbb{P}[T_{\phi_A,2}]} \frac{\partial^2 F_{|\varphi_1|,|\varphi_2| | T_{\phi_A,2}}(|\varphi_1^0|,|\varphi_2^0|) }{\partial |\varphi_1^0| \partial |\varphi_2^0|},  
  \end{split}
\end{equation}
where $F_{|\varphi_1|,|\varphi_2| | T_{\phi_A,2}}(|\varphi_1^0|,|\varphi_2^0|)$ is obtained through \eqref{A6} for $|\varphi_1^0| < \phi_A ,|\varphi_2^0|< \phi_A$ and  $\mathbb{P}[T_{\phi_A,2}]   = \frac{1 - e^{-\lambda R_L^2 \phi_A} - \lambda R_L^2 \phi_A  e^{-\lambda R_L^2 \phi_A}}{1-\frac{\Gamma(2,\lambda \pi R_L^2)}{\Gamma(2)}}$. Presenting \eqref{D1} through \eqref{eq5},  
\begin{equation} \label{D2} 
\begin{split}
&f_{|\varphi_1|,|\varphi_2| | T_{\phi_A,2}}(|\varphi_1|,|\varphi_2| )  =\frac{f_{|\varphi_1|,|\varphi_2|}(|\varphi_1|,|\varphi_2|\,, T_{\phi_A,2}) }{\mathbb{P}[T_{\phi_A,2}]} \\
 &= \frac{(\lambda_{bs}  R_L^2)^2  e^{-\lambda_{bs}  R_L^2 (|\varphi_2|-\pi)}}{e^{\lambda_{bs} \pi R_L^2 } - \lambda_{bs} R_L^2 \phi_A  e^{\lambda_{bs} R_L^2 (\pi-\phi_A)}- e^{\lambda_{bs} R_L^2 (\pi-\phi_A)}},
 \end{split}
\end{equation}
for $|\varphi_1| \in  [0,\phi_A]$ and $|\varphi_2| \in [|\varphi_1|,\phi_A]$. Now, notice that $g(|\varphi_1|), g(|\varphi_2|)$ are functions of a random variable, i.e., $g(|\varphi_i|)= g_{max} 10^{-\frac{10}{3} \Big(\frac{2 |\varphi_i|}{\phi_{3dB}}\Big)^2}$, $i \in \{1,2\}$. Building on $|\varphi_1|, |\varphi_2|$ and applying successive change of variables and simplifications, the joint PDF $f_{g(|\varphi_1|), g(|\varphi_2|) | T_{\phi_A,2}}(g_1,g_2)$ is 
\begin{equation} \label{D3} 
\begin{split}
&f_{g(|\varphi_1|), g(|\varphi_2| | T_{\phi_A,2})}(g_1,g_2) \\
&= \frac{5\,( \lambda_{bs} R_L^2 \phi_{3dB})^2}{24\,  \,g_1\, g_2} \frac{\sqrt{{\rm{log}}_{10}(\frac{g_{max}}{g_2}) {\rm{log}}_{10}(\frac{g_{max}}{g_1})}}{{\rm{ln}}(\frac{g_{max}}{g_2}) {\rm{ln}}(\frac{g_{max}}{g_1})}\\
& \times \frac{{\rm{exp}}\big(\lambda_{bs} R_L^2 \phi_{3dB} \sqrt{10 {\rm{log}}_{10}(g_{max}/g_2)}/(2 \sqrt{3})\big)}{1 - e^{-\lambda_{bs} R_L^2 \phi_A} - \lambda_{bs} R_L^2\phi_A  e^{-\lambda_{bs} R_L^2 \phi_A}},  
\end{split}
\end{equation}
for $g(|\varphi_1|) \in [g_s, g_{max}]$ and $g(|\varphi_2|) \in [g_s, g(|\varphi_1|)]$. Let $g(|\varphi_1|)=g_1$ and  $g(|\varphi_2|)=g_2$. Through $G=\frac{g_1}{g_2}$, it comes that $g_1 = G g_2$, where $g_s \leq g_1 \leq g_{max}$. Therefore,  $g_2 \in [g_s, g_{max}/z]$. Then, the PDF of $G$ can be expressed as $f_{G| T_{\phi_A,2}}(g) = \int_{g_s}^\frac{g_{max}}{g} g_2\,  f_{g(|\varphi_1|), g(|\varphi_2|) | T_{\phi_A,2}}(g\,g_2,g_2) \,{\rm d}  g_2$, for $g \in [1, g_{max}/g_s]$. After some simplifications (39) yields.

\end{document}